\documentclass[reprint,showpacs,aps,prb]{revtex4-1}
\usepackage[english]{babel}
\usepackage{amsmath,amssymb}
\usepackage{graphicx}
\usepackage{physics}
\usepackage{color}
\usepackage{slashed}
\usepackage{array}   
\newcolumntype{L}{>{$}l<{$}} 

\usepackage[section]{placeins}
\DeclareMathOperator{\E}{e}
\DeclareMathOperator{\I}{i}
\DeclareMathOperator{\V}{V}
\newcommand{\vc}[1]{{\bf#1}}
\newcommand{\dfreq}[1]{\check{#1}}
\newcommand{\bV}{\beta \V}
\usepackage{relsize}
\usepackage{soul}

\newcommand{\tst}{\textstyle}

\newcommand{\sfrac}[2]{{\tst \frac{#1}{#2}}}
\newcommand{\intsp}[1]{\sum_\sigma\int\sfrac{\dd^3{#1}}{(2\pi)^3}}\,

\newcommand{\intvp}[1]{\int\frac{\dd^2{\vc #1}}{(2\pi)^2}\,}

\newcommand{\beq}{\begin{equation}}
\newcommand{\eeq}{\end{equation}}
\newcommand{\cA}{{\mathcal A}}
\newcommand{\cI}{{\mathcal I}}
\newcommand{\vphi}{\varphi}
\newcommand{\bC}{{\mathbb C}}
\newcommand{\bI}{{\mathbb I}}
\newcommand{\bL}{{\mathbb L}}
\newcommand{\bM}{{\mathbb M}}
\newcommand{\bQ}{{\mathbb Q}}
\newcommand{\bR}{{\mathbb R}}

\usepackage[dvipsnames,table]{xcolor}
\definecolor{lightgray}{RGB}{190,190,190}
\definecolor{slategray}{RGB}{112,138,144}
\definecolor{firebrick}{RGB}{178,34,34}
\definecolor{darkorange}{RGB}{255,140,0}
\definecolor{darkgreen}{RGB}{0,100,0}
\definecolor{seagreen}{RGB}{46,139,87}
\definecolor{lightseagreen}{RGB}{32,178,170}
\definecolor{forestgreen}{RGB}{34,139,34}
\definecolor{midnightblue}{RGB}{25,25,112}
\definecolor{navyblue}{RGB}{0,0,128}
\definecolor{cornflowerblue}{RGB}{100,149,237}
\definecolor{mediumblue}{RGB}{0,0,205}

\newcommand{\neu}[1]{#1}
\newcommand{\neuneu}[1]{#1}
\newcommand{\wegdamit}[1]{}

\newcommand{\reftitel}[1]{{\sl #1}}

\def\pr#1{{#1}}

\usepackage[]{hyperref}

\begin{document}
\raggedbottom

\title{Low-Energy Effective Theory at a Quantum Critical Point of the Two-Dimensional Hubbard Model: Mean-Field Analysis}

\author{Kambis Veschgini}
\email{k.veschgini@thphys.uni-heidelberg.de}
\author{Manfred Salmhofer}
\email{m.salmhofer@thphys.uni-heidelberg.de}
\affiliation{
	Institut f\" ur Theoretische Physik, Universit\" at Heidelberg, Philosophenweg 19, 69120 Heidelberg, Germany}
\date{\today}

\begin{abstract}
\noindent
We complement previous functional renormalization group (fRG) studies of the two-dimensional Hubbard model by mean-field calculations. The focus falls on Van Hove filling and the hopping amplitude $t'/t=0.341$. The fRG data suggest a quantum critical point (QCP) in this region and in its vicinity a singular
fermionic self-energy $\Im \Sigma(\omega)/\omega \sim - \abs{\omega}^{-\gamma}$ with $\gamma\approx 0.26$ \cite{UBHD-67255998,MS-GiSa}.
Here we start a more detailed investigation of this QCP using a bosonic formulation for the effective action, where the bosons couple to the order parameter fields. To this end, we use the channel decomposition of the fermionic effective action developed in [\onlinecite{HuSa}], which allows to perform Hubbard-Stratonovich transformations for all relevant order parameter fields at any given energy scale $\Omega$.
We stop the flow at a scale $\Omega$ where the correlations of the order parameter field are already pronounced, but the flow is still regular, and derive the effective boson theory. It contains $d$-wave superconducting, magnetic, and density-density interactions. We analyze the resulting phase diagram in the mean-field approximation. We show that the singular fermionic self-energy suppresses gap formation both in the superconducting and magnetic channel already at the mean-field level, thus rounding a first-order transition (without self-energy) to a quantum phase transition (with self-energy). We give a simple effective model that shows the generality of this effect. In the two-dimensional Hubbard model, the effective density-density interaction is peaked at a nonzero frequency, so that solving the mean-field equations already involves a functional equation instead of simply a matrix equation (on a technical level, similar to incommensurate phases). Within a certain approximation, we show that such an interaction leads to a short quasiparticle lifetime.
\end{abstract}

\maketitle

\section{Introduction}

\noindent
The discovery of high-temperature superconductors and other materials displaying anomalous properties at temperatures above the transition to the symmetry-broken state, started a discussion about the breakdown of Landau's Fermi liquid (FL) behaviour. A finer characterization of what is broadly called a ``non-Fermi-liquid'' has been approached in various ways. In cases where it still makes sense to define a fermionic self-energy via a Dyson relation, the regularity properties of the self-energy hold the key to essential properties of the (quasi)-particle excitations of the many-body system. Sufficient regularity of the self-energy implies FL behaviour. A singularity of the fermionic self-energy at zero frequency implies deviations from FL behaviour. A true singularity would occur only at zero temperature, but it leaves its vestiges at positive temperature, specifically by a  small-frequency behaviour of the type $|\omega|^\alpha $ sgn$(\omega) $ with $\alpha < 1$ for $\omega \to 0$. The exponent $\alpha$ determines the anomalous exponents of the decay of the single-quasiparticle excitations, hence provides specific information about deviations from FL properties. 

These singular self-energies are also closely tied to certain quantum critical phenomena. While most often discussed in terms of the theory of an effective bosonic field, which describes the order parameter and its fluctuations, a fermionic description can in some cases be useful, in particular close to the critical point, where the order parameters vanish. 

Of the different situations where FL behaviour is expected to break down in lattice systems, we investigate here the case where the band function(s) have saddle points or even degenerate critical points, {\em Van Hove points}, on the Fermi surface, so that the latter becomes a singular Fermi surface. At these points, the fermionic density of states diverges. Consequently, the critical scales for ordering tendencies are enhanced already on the mean-field level, and in corresponding single-channel resummations of perturbation theory for the fermionic four-point function.  

On the other hand, singular features can also appear in the fermionic self-energy. A general mechanism for this was identified in \cite{MS-SFS2}, where an asymmetry in the regularity of the self-energy as a function of spatial momentum $\vc{k}$ and Matsubara frequency $k_0$ was shown to exist. For Fermi surfaces containing Van Hove points, it was proven to all orders in renormalized perturbation theory that the self-energy $\Sigma$ is at least once continuously differentiable as a function of  $\vc{k}$, and that in two dimensions, the $k_0$ derivative diverges already in second-order perturbation theory at the Van Hove points. This asymmetry is of fundamental interest, because it is directly relevant to the question of non-Fermi-liquid behaviour, and because it is a feature not present in the one-dimensional Luttinger liquids, where frequency and momentum derivatives have the same singularities in perturbation theory.

In the two-dimensional case, there are several  potential singularities in the four-point function, which are all of the same order of magnitude in perturbation theory, and which also have a nontrivial interplay with the singular terms in the self-energy. We would like to determine what really happens in this coupled system, taking into account all competing terms, and going beyond perturbation theory. The renormalization group (RG) allows us to do exactly this for weakly coupled models. 

Because of its role in modelling cuprate materials and its intrinsic interest as a prototypical, hard, model case, we study here the two-dimensional Hubbard model at Van Hove filling. Specifically, we consider the model with nearest-neighbour hopping amplitude $t>0$ and a next-to-nearest neighbour hopping amplitude $-t' < 0$, and the usual on-site repulsion $U > 0$. We  vary $\theta = t'/t$, keeping the density fixed at Van Hove filling.  The logarithmically divergent density of states implies that at temperature $\beta$, in second-order perturbation theory, the superconducting pairing term is of order $(U \log \beta)^2$, the magnetic term is of order $U \log \beta$, and the frequency derivative of the fermionic self-energy also grows like $(U \log \beta)^2$. Thus these terms can all compete with each other, and varying $\theta$ changes their relative strength. At small $\theta$, the leading correlations are antiferromagnetic. When $\theta$ gets larger, the Fermi surface is curved, hence non-nested, away from the VH points. This weakens  antiferromagnetic correlations, and the superconducting correlations dominate. At still larger $\theta$, above $0.35$, ferromagnetism dominates. 

We first observed,\cite{HSPRL,HSTflo} using the temperature-flow RG, that there is an effective cancellation of the ferromagnetic and the superconducting singularities at $\theta \approx 0.33$. Close to this value, the flow could be extended to very low scales, with numerical accuracy to zero scale. In Refs.\ \onlinecite{HSPRL,HSTflo}, the self-energy was not taken into account; nevertheless, it clearly indicated a quantum critical point separating a d-wave superconducting and a ferromagnetic phase. In the light of the above discussion, this point is of particular interest, because it can be seen already in second-order perturbation theory that two of the three competing terms cancel out. In subsequent RG studies, we used the $\Omega$-flow scheme and the vertex parametrization introduced in Ref.\ \onlinecite{HuSa}, and included a momentum- and frequency-dependent fermionic self-energy in the flow. We confirmed\cite{MS-HuGiSa,MS-GiSa} that the cancellation makes the self-energy term dominate the flow and further suppress all ordering tendencies, resulting in a quantum phase transition point at $\theta^* = 0.341$, and a non-Fermi liquid exponent $\alpha = 0.74$ in the frequency dependence of the self-energy. 

In this paper we complement our RG studies by an analysis of the order parameters, based on the results obtained by the RG. The RG flow can be stopped at any scale, yielding an effective action for the low-energy degrees of freedom. Because there are several competing terms, the interaction term has no simple factorization properties, but as shown in \onlinecite{HuSa}, it can be approximated well by a sum of boson exchange interactions, which correspond to density-density, spin-spin, and Cooper pair interactions. The parametrization of this effective action in Ref.\ \onlinecite{HuSa} is designed so that a Hubbard-Stratonovich transformation can be applied to it directly, leaving a coupled system of bosonic order parameter fields. The channel-decoupling ansatz has been refined and developed further and applied to multiband systems; see Ref.\ \onlinecite{TUF-RG} and references therein.

Here, we start our analysis of the effective bosonic system at the QCP by studying the order parameters in mean-field theory. This is only a first step in understanding the low-energy behaviour, but, due to its simplicity the method brings out an interesting aspect very clearly: the singular fermionic self-energy causes a rounding of the phase transition already in mean-field theory. This effect is explained in Section \ref{sec:mf-toy-model}. In Section \ref{sec:frgmf}, we first give details on how we use the one-particle irreducible effective action at a scale $\Omega > 0$ to define the low-energy theory, and then do the Hubbard-Stratonovich transformation and study the  mean-field theory for this low-energy theory. 
A particular, important feature of the effective action in the Hubbard model ist that it not just contains magnetic and superconducting terms, but also a density-density interaction. The latter has a rather nontrivial frequency dependence which leads to effects reminiscent of noncommensurate phases, in that the mean-field equations do not close in a subspace of small dimension. We calculate the effect of these interactions approximately in Section \ref{sec:mf-dd}.

\section{Rounding of phase transitions by singular fermionic self-energies}
\label{sec:mf-toy-model}
\noindent
In this section, we show in a model case that a singular self-energy can suppress order parameters already on the mean-field level, so that the transition from one to the other becomes continuous. The idea behind this is very simple: symmetry breaking at low temperatures and at arbitrarily low coupling strength can occur because the standard free-fermion propagator is not square-integrable at zero temperature (in fermionic models, the $L^2$ norm of the propagator is identical to the `particle-particle bubble', that is, value of the lowest-order Cooper pairing process between particles with momenta $k$ and $-k$). If, however, the denominator of the propagator contains a self-energy that vanishes like a power less than 1 as the frequency goes to zero, this term dominates the low-energy behaviour of the propagator. It makes the propagator less singular, so that the latter becomes square-integrable, and then mean-field equations no longer have solutions below a certain threshold for the coupling constant. 

In our analysis of the two-dimensional Hubbard model done in later sections of this paper,  we have isolated the fermionic self-energy as the main ingredient responsible for the quantum critical behavior. 
\neu{
At Van Hove filling and in the vicinity of the hopping parameter $t'/t=\theta^\star=0.341$,
fRG calculations predict a self-energy of the form $\propto - \I \, \mathrm{sgn}(\omega)|\omega|^{\alpha}$ with an exponent $\alpha\approx 0.74$\cite{MS-GiSa}. The dominant instability for $t'/t<\theta$ is in the \neuneu{Cooper} and for $t'/t>\theta$ in the ferromagnetic channel. We will show that mean-field calculations based on the critical self-energy correctly classify this boundary as a quantum critical point.}
In a nutshell, we will show that there is a strong connection between quantum fluctuations at the phase transition between the superconducting and the ferromagnetic phase, and the frequency dependence of the critical self-energy \wegdamit{$\propto - \I \, \mathrm{sgn}(\omega)|\omega|^{\alpha}$ with an exponent $\alpha< 1$}. The shape of the free-energy is depicted below as a function of one of the order parameters $\Delta_\text{FM/dSC}$ when the other is zero.
We note that coexistence of singlet superconductivity and ferromagnetism is excluded at mean-field level because the corresponding stationary point is a maximum of the free energy. Triplet pairing is suppressed in the Hubbard model at Van Hove filling because the form factor of the gap must be odd, which implies that it also vanishes at the boundary of the first Brillouin zone, hence at the saddle points of the dispersion relation. 
\begin{center}
    \includegraphics{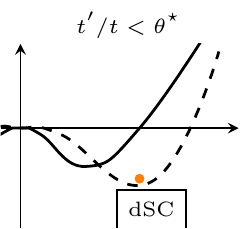}
    \includegraphics{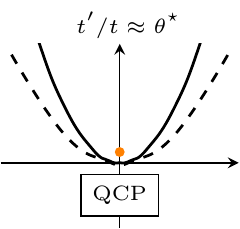}
    \includegraphics{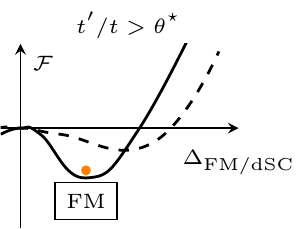}
\end{center}
In the second order quantum phase transition, both order parameters vanish in the vicinity of the critical point. In the following we will demonstrate that the critical exponent of the self-energy is crucial for second order quantum phase transition. 

~\\
Consider the mean-field action
\begin{equation}
	\label{eq:toy-mode-action}
	\begin{split}
	S =& -\intsp{k}(\I k_0 -\epsilon_{\vc k}-\Sigma(k_0)) {\overline\psi}_{k,\sigma} \psi_{k,\sigma} \\
	&+  \Delta_\text{x} \cdot \widehat O_\text{x}+\beta V \frac{\abs{\Delta_\text{x}}^2}{U_\text{x}}\;,
	\end{split}
\end{equation}
where $k=(k_0,k_1,k_2) = (k_0, {\vc k})$ contains the Matsubara frequency variable $k_0$ and spatial momentum ${\vc k} = (k_1,k_2)$. 
$\epsilon_{\vc k}$ is the dispersion relation
\begin{equation}
	\epsilon_{\vc k} = -2 t \qty(\cos k_1+\cos k_2) + 4 t' \qty(\cos k_1\, \cos k_2 + 1) \;
\end{equation}
of the two dimensional Hubbard model at Van Hove filling. $\Sigma$ is the self-energy and
\begin{align}
    \widehat O_{\text{FM}} &= \intsp{k} \sigma\,{\overline\psi}_{k,\sigma}\psi_{k,\sigma}\;,\\
    \widehat O_{\text{dSC}} &= \frac12 \intsp{k} \sigma\,f_2(\vc k){\overline\psi}_{k,\sigma}{\overline\psi}_{-k,-\sigma}\;.
\end{align}
are operators quadratic in the fields $\psi,{\overline\psi}$ whose expectation values $O_\text{x}= \langle \widehat O_\text{x}\rangle$, $\text{x}\in\{\text{FM},\text{dSC}\}$ are the order parameter of the ferromagnetic and superconducting states.
We will use a self-energy of the form
\begin{equation} \label{eq:toy-model-se}
	\Sigma(\omega) \propto - \I \, \mathrm{sgn}(\omega)|\omega|^{1-\gamma} \;,\;\gamma=0.26
\end{equation}
motivated by fRG calculations \cite{MS-GiSa}.
More precisely we restrict this self-energy to the vicinity of the critical hopping $ \theta^\star=0.341$ through
\begin{equation} \label{eq:toy-model-se-theta}
	\begin{split}
	\Sigma(\omega) = -\I \text{sgn}(\omega)\Big[& \qty(\omega^2+(t'/t-\theta^\star)^2)^{\frac{1-\gamma}{2}}\\
	& - \qty|t'/t-\theta^\star|^{1-\gamma} \Big]\;.
	\end{split}
\end{equation}

The gap parameters $\Delta_\text{X}$ arise as the zero modes of Hubbard-Stratonovich fields. $U_\text{X}>0$ denotes the corresponding effective interaction strength. Note that $U_\text{dSC}>0$ corresponds to an attractive effective interaction which gets generated during the fRG flow.

Define $F_\Sigma$ by 
\begin{equation}
	F_\Sigma(\xi) = \int \frac{\dd{\omega}}{2\pi}\, \frac{1}{\I(\omega-\Im \Sigma(\omega))-\xi}\quad
\end{equation}
The self-consistency equations for the gap parameters, minimizing the free-energy,
 are given by
\begin{subequations}
\begin{align}
	1 =& \frac{U_\text{FM}}{2} \intvp{k} \frac{\sigma\,F_\Sigma(\epsilon_{\vc k}-\sigma \Delta_\text{FM})}{\Delta_\text{FM}}\;, \label{eq:toy-mode-fm-fermi}\\
	1 = & \frac{U_\text{dSC}}{2} \intvp{k} \frac{f_2^2(\vc k)}{E_{\vc k}}\sigma F_\Sigma\qty(-\sigma E_{\vc k})\;,  \label{eq:toy-mode-dsc-fermi}
\end{align}
\end{subequations}
with
\begin{equation}
	E_{\vc k} = \sqrt{\epsilon_{\vc k}^2+f_2^2(\vc k)\abs{\Delta_\text{dSC}}^2}\;.
\end{equation}
and the order parameters are given by $O_\text{X} = 2 \Delta_\text{X} / U_\text{X}$.

\begin{figure}[]
    \includegraphics{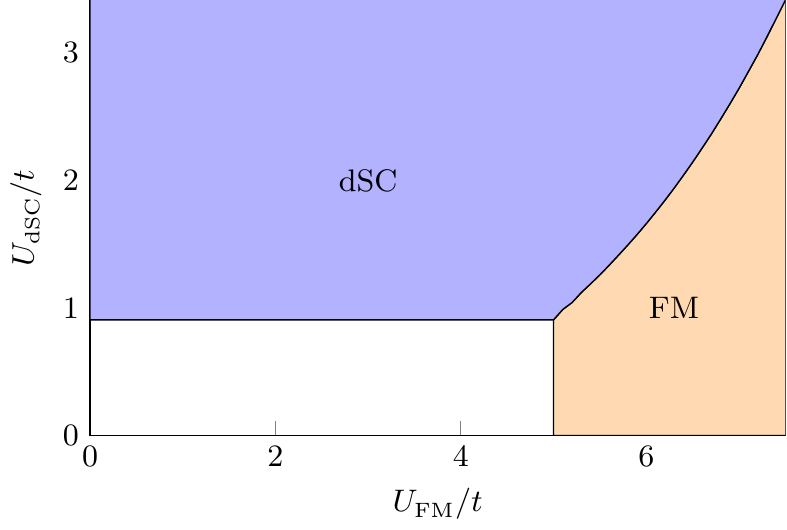}

	\caption[]{Zero temperature phase diagram at $t'/t=\theta^\star$.  Due to the critical self-energy there is no symmetry breaking in the unshaded region $U_\text{FM}/t<5,U_\text{dSC}/t<0.9$. }
	\label{fig:toy-model-transition}
\end{figure}

We denote the smallest interaction $U_\text{X}$ from where on the corresponding self-consistency equation has a solution ($\Delta_\text{X}\neq 0$) with $U_\text{X}^\text{min}$,
\begin{equation}
\label{eq:toy-model-lower-bounds}
\begin{split}
	U_\text{FM}^\text{min}/t &= \qty(-\intvp{k} F_\Sigma'(\epsilon_{\vc k}))^{-1}\;,\\
	U_\text{dSC}^\text{min}/t &= \qty(\frac12 \intvp{k} \frac{f_2^2(\vc k)}{|\epsilon_{\vc k}|}\qty(F_\Sigma\qty(-|\epsilon_{\vc k}|)-F_\Sigma\qty(|\epsilon_{\vc k}|)))^{-1}\;.
\end{split}
\end{equation}
If we neglect the self-energy then, at Van Hove filling, $U_\text{FM}^\text{min}=U_\text{dSC}^\text{min}=0$. That is, there is no threshold below which either of the two orderings becomes impossible. Depending on the free energy, the ground state will either be superconducting or ferromagnetic, and the transition from one to the other is discontinuous. Including the self-energy eq.~\eqref{eq:toy-model-se} changes this picture drastically: in this case, $U_\text{FM}^\text{min},U_\text{dSC}^\text{min}>0$, so that a quantum critical regime becomes possible. The \neu{phase diagram} for $t'/t=\theta^\star$ is shown in Fig.~\ref{fig:toy-model-transition}. \neu{For small enough effective interaction the system will be in a quantum critical regime}. 
\wegdamit{In the next section, where we include the scale dependencies, $U_\text{FM}^\text{min},U_\text{dSC}^\text{min}$ will grow strongly during the flow. Starting with the bare interaction $U/t=3$, we stop the flow when the effective interaction gets as large as $20t$. Yet, $U_\text{FM/dSC}^\text{min}$ also grow as the scale drops, so that ordered phases remain suppressed close to $\theta^\star$.}

\begin{figure}[]
    \includegraphics{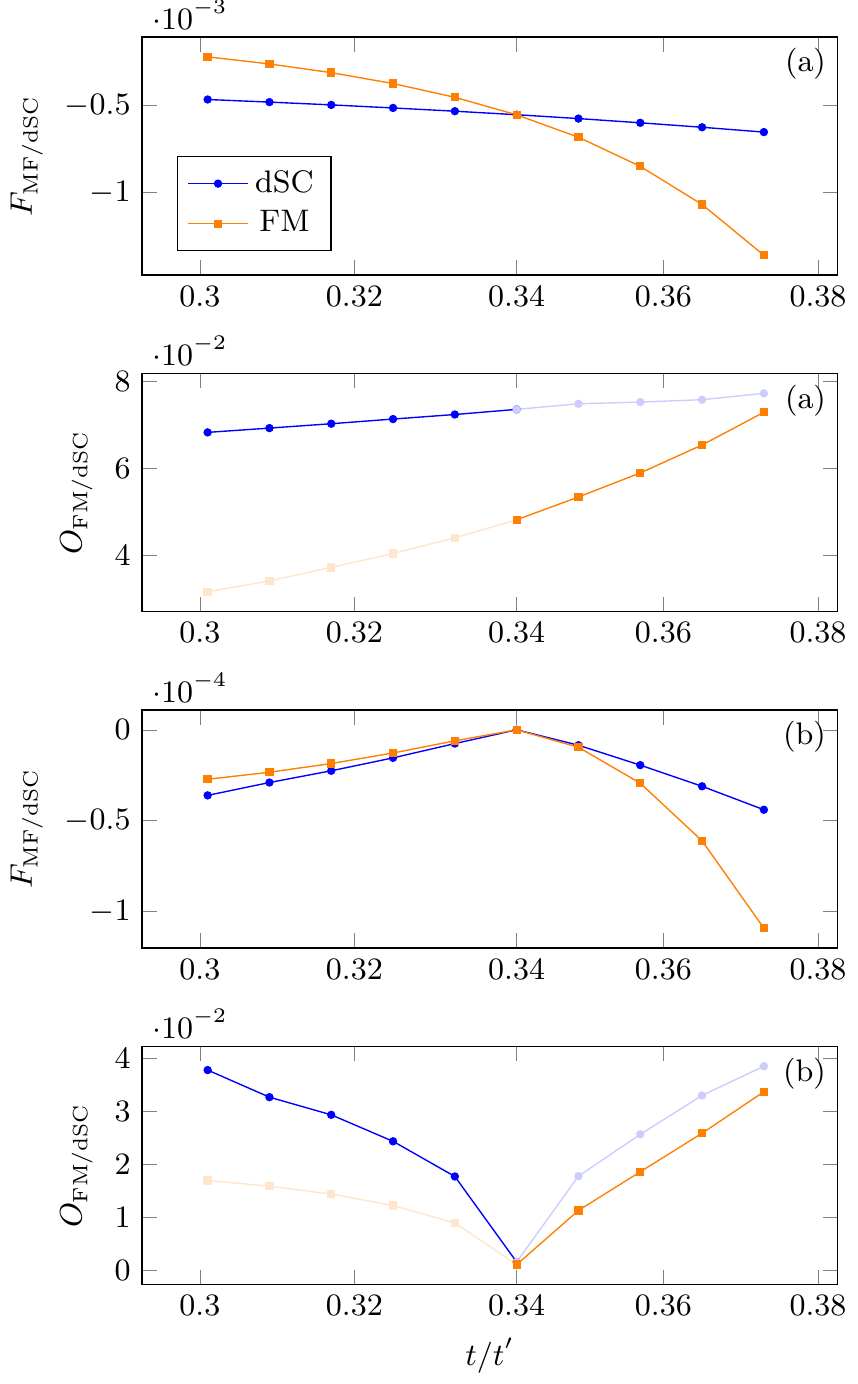}

	\caption[]{Free energy and order parameters as functions of the hopping amplitude $t'/t$. (a) Without self-energy, $U_\text{FM}/t=2.61$ and $U_\text{dSC}/t=0.42$ (b) With critical self-energy, $U_\text{FM}/t=4$ and $U_\text{dSC}/t=0.42$. Lighter color indicates that the order parameter corresponds to the phase with the higher free energy. The interactions are chosen such that the phase transition happens close to $t'/t=\theta^\star$.}
	\label{fig:toy-model-theta}
\end{figure}

Finally in Figure \ref{fig:toy-model-theta} we show the order parameters as a function of the hopping amplitude with and without the self-energy. With the self-energy we see a quantum critical region. Close to the QCP the free-energy as a function of $\Delta_\text{FM}$ and $\Delta_\text{dSC}$ has only a trivial minimum at $\Delta_\text{FM} = \Delta_\text{dSC} = 0$. In our convention, in such a case $F_\text{FM} = F_\text{dSc} = 0$.  Neglecting the self-energy results in non-vanishing order parameters. Each one corresponds to a non-trivial minimum of the free energy. The deepest minimum determines the state. At $\theta^\star$ there is a first order transition from a superconductor to a ferromagnet since $F_\text{FM}$ becomes smaller than $F_\text{dSC}$ as we move from the left to the right along the $t'/t$-axis. 

The numerical values for the parameters that we have used in this model case serve the purpose of demonstrating the basic effects of the self-energy on gap formation.
While qualitatively similar, the full model contains nontrivial dependencies of all parameters on the hopping amplitude and scale. Furthermore, not only the self-energy, but also the interaction depends on momenta and frequencies.  Nevertheless, the frequency dependence of the self-energy remains the driving force suppressing the gap formation in the vicinity of the transition point. In the next section we will show this using our quantitative calculations for the two-dimensional Hubbard model. We discuss the low-energy obtained from the RG when the flow is stopped at a certain scale, and then evaluate the remaining functional integral for the partition function in a saddle-point approximation.

\section{The low-energy effective theory obtained from the renormalization group}
\label{sec:frgmf}
\noindent
An essential feature of the RG approach is that the scale parameter, which in our case is an energy scale $\Omega$, not only determines how degrees of freedom are successively integrated over, but also allows to use the effective action obtained at a certain energy scale $\Omega$ to define the low-energy theory. 

In this way, one can then use a variety of methods at lower scales. This is in many cases of great practical interest because the continuation of a flow with several competing order parameters into symmetry-broken phases is involved, and it is useful to get information by simplified means before embarking on a full analysis. 

\subsection{General structure}
\noindent
We briefly recapitulate how the effective action obtained at a certain energy scale $\Omega$ from the fermionic renormalization group flow is used to define a low-energy model.  

The generating functional for the microscopic theory is given by 
\beq
Z (\bar\eta, \eta)
=
\mathcal{N} \int \mathrm{d}\mu_{C}(\bar \phi,\phi) \E^{-\cI_0(\bar \phi,\phi)}\;
  \E^{(\bar\eta,\phi) - (\bar \phi,\eta)}
\eeq
where $\mu_{C}$ denotes the normalized Grassmann Gaussian measure
\beq
\mathrm{d}\mu_{C} 
= 
\det C\; 
\prod_{k,\sigma} \dd \bar \psi_{k,\sigma} \dd \psi_{k,\sigma}\;\E^{(\bar \psi, C^{-1} \psi)}
\eeq
and the bare propagator $C$ is determined by the quadratic part of the Hamiltonian and $\cI_0$ by the interaction terms. The normalization constant $\mathcal{N}$  is simply the partition function for free fermions:
$ \mathcal{N} =  \prod_{\sigma,{\vc k}} \beta V (1+ \E^{-\beta \epsilon_{\vc k}}) =\prod_{\sigma}\prod_{p}\beta V C(k_0,\vc k)$. $V=L^2$ is the surface area of the lattice and $\beta$ is the inverse temperature. Our results correspond to the limit $\V\to\infty$ and $\beta \to \infty$. In our fermionic model, we assume that $\cI_0$ does not contain any terms with odd powers in the fields. The RG method starts by decomposing 
$C = C_\Omega+ D_\Omega$, where 
\beq
C_\Omega = C \chi_\Omega, 
\qquad
D_\Omega = C - C_\Omega = C (1- \chi_\Omega) \; .
\eeq
Here $\chi_\Omega$ is an infrared regulator that depends on the scale parameter $\Omega$.
In our case, 
\beq\label{eq:Omegareg}
C(k_0, \vc k) = \frac{1}{\I {p}_0-\xi(\vc k)}
\eeq 
times $\delta_{\sigma,\sigma'}$, where $\sigma$ denotes the spin index, and we choose
\beq
\chi_\Omega (k_0, \vc k)
=
\frac{k_0^2}{k_0^2 + \Omega^2}
\eeq
independent of $\vc k$. $\chi_\Omega$ is a regulator because the $k_0^2$ in the numerator cancels the singularity of $C$ at $k_0=0$. This soft regulator is chosen\cite{HuSa} to avoid artificial suppression of ferromagnetism and other small-momentum correlations. Moreover, including the Fermi surface deformation does not require any adaptive scale decomposition \cite{dynRG} with this regulator because $\chi_\Omega$ is independent of $\vc k$. The low-energy propagator
\beq
D_\Omega (k_0,\vc k)
=
\frac{\Omega^2}{k_0^2 + \Omega^2}
\;
\frac{1}{\I {p}_0-\xi(\vc k)}
\eeq
has the same singularity at $k_0 =0$ as $C_\Omega$, but it decays as $|k_0|^{-3}$ for $|k_0| \to \infty$. 

The following considerations about the effective theory below scale $\Omega$ apply in general and do not depend on our particular choice of model and regulator. It follows by the addition principle of Gaussian integration \cite{msbook} that the integration field splits as $\phi = \psi + \vphi$ into a `high-energy' field $\vphi$ and a `low-energy' field $\psi$, and
\beq
\begin{split}
Z (\bar\eta, \eta)
= 
&\mathcal{N}\int \mathrm{d}\mu_{D_\Omega}(\bar \psi,\psi) \E^{(\bar\eta,\psi) - (\bar \psi,\eta)}\\
&\mathcal{N}\int \mathrm{d}\mu_{C_\Omega}(\bar \vphi,\vphi) \E^{-\cI_0(\bar \psi + \bar\vphi,\psi+\vphi)}
\E^{(\bar\eta,\vphi) - (\bar\vphi,\eta)}
\end{split}
\eeq
If we are interested only in the correlations of the $\psi$ and $\bar\psi$ fields at low energy scales, we may choose the sources to couple only to the low-energy fields $\psi$ and $\bar\psi$,\footnote{For a strict cutoff function $\chi_\Omega$, which vanishes identically in an interval of length $\Omega$, this can be achieved by choosing the support of the source fields to be in that interval. For a soft regulator, as in (\ref{eq:Omegareg}), this is an approximation} and get 
\beq\label{eq:loetheo}
Z (\bar\eta, \eta)
= 
\mathcal{N}\int \mathrm{d}\mu_{D_\Omega}(\bar \psi,\psi) \E^{(\bar\eta,\psi) - (\bar \psi,\eta)}
\;
\E^{-\cA_\Omega(\bar\psi,\psi)} \; ,
\eeq
where $\cA_\Omega(\bar\psi,\psi) = -\log( 
\mathcal{N} \int \mathrm{d}\mu_{C_\Omega}(\bar \vphi,\vphi) \E^{-\cI_0(\bar \psi + \bar\vphi,\psi+\vphi)})
$ is Wilson's effective interaction, namely the generating functional of the connected, and $C_\Omega$-amputated, correlation functions. Once $\cA$ has been obtained, \eqref{eq:loetheo} is the definition of the partition function of the `low-energy effective theory'. 

In our RG flow, we calculate the one-particle irreducible vertex functions. The interaction $\cI_\Omega$ is given in terms of these vertices in the following way. The quadratic part of $\cA$ is $(\bar\psi, A_\Omega^{(2)} \psi)$, where
\beq\label{eq:Azwo}
A_\Omega^{(2)} = C_\Omega^{-1} - C_\Omega^{-1} G_\Omega C_\Omega^{-1} \; ,
\eeq
and $G_\Omega$, the full propagator above scale $\Omega$, is related to the selfenergy $\Sigma_\Omega$  by a Dyson relation
\beq\label{eq:cutoffdyson}
G_\Omega
=
(C_\Omega^{-1} - \Sigma_\Omega)^{-1}
=
\frac{\chi_\Omega}{C - \chi_\Omega \Sigma_\Omega} 
\eeq
in which the regulator function $\chi_\Omega$ multiplies the selfenergy in the denominator.
The quartic and higher terms of $\cA$ are given by sums over tree diagrams, the lines of which carry full propagators $G_\Omega$, and the vertices of which are given by the 1PI vertices $\Gamma^{(2m)}$ with $m \ge 2$. 
It follows that external legs carry a factor $G$ times $C^{-1}$
In particular, 
\beq\label{eq:mf-a4}
A^{(4)}_\Omega(K_1,\cdots, K_4) = \Gamma^{(4)}_\Omega(K_1,\cdots,K_4)\prod_{i=1}^4 \frac{G_\Omega(k_i)}{C_\Omega(k_i)}\;,
\eeq
where $K_i = ((k_0)_i, \vc k_i,\sigma_i)$. This formula holds for the four-point function because, by our assumption that the microscopic interaction has no odd interaction terms, the connected four-point function is obtained by attaching full propagators to the irreducible four-point vertex. The denominators in this formula reflect the amputation by $C_\Omega$. In the following we drop all $A^{(2m)}_\Omega$ with $m \ge 3$. 

The form of the interaction makes it natural to change variables to the fields
\begin{equation} \label{eq:mf-new-grassmann-vars}
(\overline\Psi_{k,\sigma},\Psi_{k,\sigma}) =  \frac{G_\Omega(k)}{C_\Omega(k)} (\overline\psi_{k,\sigma},\psi_{k,\sigma})\;,
\end{equation}
so that the generating functional $Z (\bar\eta, \eta)= {\cal N}_\Omega  \tilde Z (\bar H, H)$
with 
$
{\cal N}_\Omega
=
\mathcal{N}
\det (1- C_\Omega \Sigma_\Omega)^{-1} 
$
and 
\beq \label{eq:Z-new-vars}
\begin{split}
\tilde Z (\bar H, H)
&=
\int \mathrm{d}\mu_{T_\Omega}
{\scriptstyle (\bar \Psi,\Psi)}\;
\E^{-\Gamma^{(4)}_\Omega(\bar\Psi,\Psi) + (\bar H,\Psi) - (\bar \Psi,H)}
\;
\end{split}
\eeq
is a function of the rescaled source fields
\beq
\bar H = (1-C_\Omega \Sigma_\Omega)^{-1}\bar \eta,
\quad
H = (1-C_\Omega \Sigma_\Omega)^{-1} \eta \; .
\eeq
The propagator of the $\Psi$ fields is
\beq \label{eq:def-covar-T}
T_\Omega 
=
(1-\chi_\Omega)\;  (C^{-1} - \Sigma_\Omega)^{-1} (1-C_\Omega\Sigma_\Omega)^{-1} \; .
\eeq
The first factor $1-\chi_\Omega$ in $T_\Omega$ (in our case, $\Omega^2/(\Omega^2+k_0^2)$) now suppresses large energies, as is appropriate for a low-energy theory. The factor in the middle
is a full propagator, in which the selfenergy of the fields that were integrated over enters in the standard way, but, in contrast to (\ref{eq:cutoffdyson}), {\em without} getting multiplied by a regulator function. The last factor is there because the quadratic term of the effective action $\cA$ is reducible, hence may contain strings of self-energy insertions from the integration of scales above $\Omega$. 
The condition that this factor is nonsingular poses a restriction on the size of the effective interaction, hence, in flows where the vertex functions grow, it also provides a test whether 
the flow equation still makes sense. 

\subsection{The effective interaction of the Hubbard model}

\noindent
The above setup does not allow us to continue the flow into the symmetry-broken phase, mainly because we have made a symmetric ansatz for the effective action. The fermionic RG flow can be continued into symmetry-broken phases by including a small symmetry-breaking field, so that nonvanishing expectation values of order parameter fields can develop, and then turning the symmetry-breaking field to zero after the limit $\Omega \to 0$ has been taken \cite{SHML,gersch08,eberlein10,eberlein13}. One can also use partial bosonization to follow the flow into the symmetry broken phase \cite{aletWetterich,Wetzel}.
With our symmetric ansatz, the flow runs into a singularity at some positive $\Omega_s$, which implies that we have to restrict the flow to scales above some $\Omega^* > \Omega_s$, where the coupling functions are still finite, and not too large. The scale $\Omega_s$ gives an estimate for the critical temperature, which is usually an overestimate, because the fluctuations of the order parameter fields can further suppress the order parameters. (In two dimensions, it is this suppression that yields the Mermin-Wagner theorem.) At a `deconfined' QCP that is not shielded by some ordered phase, $\Omega_s =0$, so that the flow can be taken to zero. In principle, this provides a way to test for the existence of such a deconfined QCP. In our model, we can run the flow to scales as low as $\Omega/t\sim 10^{-5}$, but lower scales are hard to access because the accurate evaluation of bubble integrals becomes challenging. Thus we stop the flow at a low scale, bosonize the effective interaction obtained from the RG flow, and study the remaining nontrivial functional integration over the low-energy degrees of freedom in a bosonic language. In this paper, we apply a saddle-point approximation for the bosonic integration, which corresponds to mean-field theory for the order parameters. This procedure also gives information about the order parameters. 

To simplify notation, we drop the subscript $\Omega$ whenever the scale dependence is clear from the definition of a quantity. In particular, we denote the effective two-particle interaction at scale $\Omega$, $\Gamma^{(4)}_\Omega$, by $V$ in the following. 
The general $SU(2) \times U(1)$-symmetric form of $V$ is 
\begin{equation}\label{eq:full-interaction}
\begin{split}
V(\bar\psi,\psi)  
= 
\frac12 \int \prod_{j=1}^4 
\sfrac{\dd^3 p_j}{(2\pi)^3} \;  
&\delta(p_2+p_2-p_3-p_4) 
\\ 
\times \; v(p_1,p_2,p_3)
& \sum_{\sigma,\tau}
\bar \psi_{\sigma\,p_1}\bar \psi_{\tau\,p_2} \psi_{\tau\,p_3} \psi_{\sigma\,p_4}\;.
\end{split}
\end{equation}
To capture the singular momentum dependence of the vertex in an efficient parametrization
the interaction vertex (\ref{eq:full-interaction}) is decomposed into different
channels\cite{HuSa},
\begin{equation} \label{eq:channel-splitting}
V =  V_B +  V_K +  V_M +  V_D \;,
\end{equation}
as follows. $V_B$ is the bare Hubbard interaction,
\begin{equation} \label{eq:channel-splitting2}
\begin{split}
V_K(\overline\psi,\psi) 
&= 
-\sfrac14 \int\sfrac{\dd^3{\ell}}{(2\pi)^3} \sum_{m,n=1}^\infty K_{m,n}(\ell)\, S^{(0)}_m(\ell) S^{(0)}_n(-\ell),
\\
V_M(\overline\psi,\psi) 
&= 
-\sfrac14 \int\sfrac{\dd^3{\ell}}{(2\pi)^3} \sum_{m,n=1}^\infty M_{m,n}(\ell)\sum_{j=1}^3 S^{(j)}_n(\ell) S^{(j)}_m(-\ell),
\\
V_D(\overline\psi,\psi) 
&= 
+\int\sfrac{\dd^3{\ell}}{(2\pi)^3} \sum_{m,n=1}^\infty D_{m,n}(\ell)\sum_{j=0}^3 \bar C^{(j)}_m(\ell) C^{(j)}_n(\ell),
\end{split}
\end{equation}
 and \(S^{(j)}_\ell,C^{(j)}_\ell,\bar C^{(j)}_\ell\) denote the
fermionic bilinears
\begin{equation}\label{eq:bilinears-def}
\begin{split}
S^{(j)}_m(\ell) &= \int\sfrac{\dd^3{q}}{(2\pi)^3} \;  f_m(\vc q) \;\overline\psi_q^T \sigma^{(j)} \psi_{q+\ell}\;,\\
\bar C^{(j)}_m(\ell) &= \frac{\I}{2} \int\sfrac{\dd^3{q}}{(2\pi)^3} \; f_m(\vc q) \; \overline\psi_q^T \sigma^{(j)} \overline\psi_{\ell-q}\;,\\
C^{(j)}_m(\ell) &= \frac{\I}{2} \int\sfrac{\dd^3{q}}{(2\pi)^3}\; f_m(\vc q)  \;\psi_q^T \sigma^{(j)} \psi_{\ell-q}\;,
\end{split}
\end{equation}
where $\psi(p) = (\psi_+(p), \psi_-(p))^T$, similarly for $\overline \psi(p)$, $\sigma^{(j)}$ are the Pauli matrices ($\sigma^{(0)} = 1$) and \(f_{m}\) are scale independent form factors, in particular 
\begin{equation}\label{formfactors}
\begin{split}
f_1(\vc q) &= 1 \;,\\
f_2(\vc q) &= \cos(q_x)-\cos(q_y)\;.
\end{split}
\end{equation} 
In the case $m=1$ we drop the subscript from the bilinear, writing $S^{(0)}$ for $S^{(0)}_1$.
The full interaction vertex is then given in terms of the bosonic
propagators \(K_{m,n}\), \(M_{m,n}\), \(D_{m,n}\) and the form factors $f_n$ by
\begin{equation}
v(p_1,p_2,p_3)
= 
U+\sum_{m,n=0}^\infty V_{m,n} (p_1,p_2,p_3)
\end{equation}
with
\begin{equation}
\begin{split}
&V_{m,n} (p_1,p_2,p_3)
\\
&=
{\tst 
\hphantom{\frac12}
f_m(p_1+\frac{p_3-p_1}{2}) \; M_{m,n}(p_3-p_1)\; f_n(p_2-\frac{p_3-p_1}{2}) }
\\
&+
{\tst 
\frac12 f_m(p_1+\frac{p_2-p_3}{2})\;  M_{m,n}(p_2-p_3)\; f_n(p_2-\frac{p_2-p_3}{2}) }
\\
&-
{\tst 
\frac12 f_m(p_1+\frac{p_2-p_3}{2})\;  K_{m,n}(p_2-p_3)\; f_n(p_2-\frac{p_2-p_3}{2})} 
\\
&-
{\tst \hphantom{\frac12}
f_m(\frac{p_1+p_2}{2}-p_1)\;  D_{m,n}(p_1+p_2)\; f_n(\frac{p_1+p_2}{2}-p_3)} \; .
\end{split}
\label{eq:v-from-decomposition}
\end{equation} 
In Eq.~\eqref{eq:channel-splitting} we have kept the initial interaction separate so that the exchange propagators are zero at the beginning of the flow.

The interpretation of this ansatz for the effective interaction is that composite operators given the fermionic bilinears, here density operators, Cooper pair fields, and spin fields, have coupling functions 
$K$, $D$, and $M$. Thus we have an easy to understand effective interaction. Moreover, there is the kinetic term of the low-energy fermionic degrees of freedom, which, importantly, contains the self-energy from the integration down to scale $\Omega$. In the general setup of Ref.\ \onlinecite{HuSa}, the indices  $m$ and $n$ in (\ref{eq:channel-splitting2}) label members of an orthonormal basis of functions on momentum space, so that, in principle, every square-integrable function can be represented this way. The main idea of this decoupling is that singularities in $v$ that develop during the RG flow are captured by the functions $D_{m,n}$, $M_{m,n}$ and $K_{m,n}$, which can be thought of as boson exchange propagators, while the form factors of the fermionic bilinears are regular functions, hence square-integrable, so that the expansion in $m$ and $n$ applies. Clearly, several remarks are in place here. First, boson propagators need to have positivity properties to preserve stability; this is, in fact, not a problem in the fermionic RG flow---rather, it provides a test whether the effective action can really be bosonized. Second, the occurrence of singularities only as functions of $p_1 - p_3$, $p_2 - p_3$, and $p_1 + p_2$ can be strictly proven for small coupling functions,\cite{SaRMP,Sa98} but in later stages of the flow it is an assumption. Third, when entering the symmetry-broken phase, where some of $D_{m,n}$, $M_{m,n}$ and $K_{m,n}$ develop singularities, the form factors themselves become singular and this channel decomposition is no longer accurate, and must be refined \cite{Eberleinfein}. We shall, however, use it only in the symmetric phase; there it is well-justified, and it has the advantage of allowing to switch to a description in terms of bosonic order parameter fields by a straightforward Hubbard-Stratonovich transformation (provided the above-mentioned positivity holds). 

In the general parametrization of Ref.\ \onlinecite{HuSa}, all functions may depend both on the spatial momenta and on the Matsubara frequencies. We make the approximation that the form factors are independent of the frequencies (see (\ref{formfactors})), but we keep the frequency dependence of the functions $D_{m,n}$, $M_{m,n}$ and $K_{m,n}$. In Ref.\ \onlinecite{HuSa,MS-HuGiSa}, we verified that this approximation, and keeping only \(K_{1,1}\), \(M_{1,1}\), \(D_{1,1}\), and \(D_{2,2}\), lead to an accurate representation of the flow in the two-dimensional square-lattice Hubbard model with a frequency-dependent self-energy. Thus we drop all other pairs $(m,n)$ from the sums in (\ref{eq:channel-splitting2}). 

The exchange propagators \(M_{1,1}\), \(D_{1,1}\) and \(D_{2,2}\) remain positive during the flow. The function \(K_{1,1}\) is positive at zero frequency, but it develops a pronounced negative minimum at nonzero frequency. At a first glance, the positive static part signals an attractive density-density interaction, but $M_{1,1}$ also contributes a term to this interaction channel. In particular we can decompose a local interaction $-\frac14 M(0)\int {\vc S}(\ell){\vc S}(-\ell) \dd{\ell}$ into a magnetic interaction $-\frac14 M(0)\int S^{(3)}(\ell)S^{(3)}(-\ell) \dd{\ell}$ and a density-density interaction $-\frac12 M(0)\int S^{(0)}(\ell)S^{(0)}(-\ell) \dd{\ell}$. For the bare model, this decomposition leads to the same mean-field equations as the generalized Hartre-Fock theory\cite{BLS,Langmann}. 

In Section \ref{sec:mf-dd}, we investigate the effects of the frequency-dependent interaction $K$ on the self-energy, which is of particular interest at the QCP. Elsewhere we assume an approximately static density. 

\subsection{The effective interaction in bosonic form}
\label{sec:bozo}
\noindent
The decomposition (\ref{eq:channel-splitting2}) is made such that  we can easily bosonize the effective action. Assuming for the moment that all functions $M$, $D$, and $K$ define positive quadratic forms, this is done by the following identities for the characteristic functions of Gaussian measures: let $Q$ be a symmetric $N \times N$ matrix, with positive definite real part Re $Q$, so that $(\phi, Q \phi) = \sum_{i,j=1}^N \phi_i Q_{i,j} \phi_j $  has positive real part for all real $N$-vectors $\phi \ne 0$, denote $R = Q^{-1}$, and let the normalized Gaussian measure $\dd\gamma_{R}(\phi) = (\det 2\pi R)^{-\frac12} \E^{- \frac12 (\phi, Q \phi)} \dd^N \phi$, then 
\beq\label{HS1}
\E^{\frac12(b, R b)}
=
\int_{\bR^N} \dd\gamma_{R}(\phi) \; \E^{ (b, \phi)} 
\eeq
holds for any $b \in \bC^N$ (and therefore also for any $b$ in the even subalgebra of a Grassmann algebra). 
More generally, if $H$ is a complex matrix and its hermitian part $\frac12(H+H^\dagger)$ is positive definite, so that for all complex $N$-vectors $\phi \ne 0$, $(\bar \phi, H \phi) = \sum_{i,j=1}^N \overline \phi_i H_{i,j} \phi_j$ has a positive real part, let the normalized complex Gaussian measure with covariance $K= H^{-1}$ be defined as $\dd\gamma_{K}(\bar\phi, \phi) = (\det \pi K)^{-1} \E^{- (\phi, H \phi)} \; \dd^N \bar \phi \; \dd^N \phi$, then 
\beq\label{HS2}
\E^{(\tilde b, K b)}
=
\int_{\bC^N} \dd\gamma_{K}(\bar \phi,\phi) \; \E^{ (b, \phi) + (\tilde b, \bar\phi)} 
\eeq
holds for all $b, \tilde b \in \bC^N$ (and hence also for any $N$-vectors $b$ and $\tilde b$ with components that are even elements of a Grassmann algebra). 

Some care is required when using complex integrals in a truly infinite-dimensional setting of a functional integral, because for complex measures the above conditions on the real (or hermitian) part do not suffice to define a sigma-additive measure. However, in the application to many-body models, the functional integral always results as a limit of a time discretization by a Trotter formula. Before the limit is taken, all these manipulations make sense, and considering normalized correlation functions allows to bypass questions about the existence of measures. 

In the representation (\ref{eq:channel-splitting2}) of the terms in the effective action (\ref{eq:channel-splitting}), the functions $K_{m,n}$, $M_{m,n}$, and $D_{m,n}$ play the role of the covariance of the bosonic Gaussian measures. For the density-density interaction $K$ and the magnetic interaction $M$, the fields can be chosen real, since they couple the same type of fermionic bilinears. The Cooper pair interaction $D$ couples a Cooper pair bilinear with its conjugate, so the corresponding boson field must be chosen complex. Since the functions are already given in diagonal representation in momentum space, checking if their real (hermitian) part is positive amounts to checking that the real part of the function is positive. This is an issue for $K$, which is not always positive, but not for $M$ and $D$. 
We will discuss this further in section \ref{sec:mf-dd}.

We note in passing that it is not necessary to choose the covariance of the boson fields exactly equal to $K$, $M$, or $D$. If convenient, we may also take only an approximation to these functions, as a matter of convenience. If, say, we take only $M_0$ instead of $M$, a fermionic four-point vertex with $m=M-M_0$ remains. If $m$ is small enough, it can be taken into account by perturbation theory. 

We can write the effective interaction (\ref{eq:channel-splitting2}) as
\beq
\begin{split}
V_K 
&= 
-\sfrac14 ( S^{(0)}, \; (K_{1,1} - 4U) \, S^{(0)})
\\
V_M
&=
-\sfrac14 \sum_{j=1}^3 
( S^{(j)}, \; M_{1,1} \, S^{(j)})
\\
V_D
&=
(\bar C_2^{(2)} , \; D_{2,2} \, C_2^{(2)})
\end{split}
\eeq
where the bilinear form is now $(f,g) = \int \sfrac{\dd^3 \ell}{(2\pi)^3} f(\ell) g(\ell)$. 
Here we have already restricted to singlet $d$-wave Cooper pairing by taking only $C_2$ in the Cooper term because the the triplet form factor vanishes at the Van Hove points, hence is irrelevant at Van Hove filling, and because the $s$-wave interaction is repulsive, hence will not lead to pairing. 

The Hubbard-Stratonovich (HS) transformation is now a straightforward application of (\ref{HS1}) and (\ref{HS2}) to the remaining Grassmann integral (\ref{eq:Z-new-vars}). It introduces three types of fields, corresponding to the density-density interaction $V_B + V_K$, the spin-spin interaction $V_M$ and the  singlet Cooper pair interaction $V_D$ in (\ref{eq:channel-splitting}). The HS field for the density interaction is a real scalar, the one for the magnetic interaction is a three-component vector field, and the one for the superconducting interaction is a complex scalar. We write the normalized expectation value with respect to the corresponding Gaussian measures as 
\beq
\begin{split}
\left\langle 
F
\right\rangle_{K,M,D}
=
&\int \dd\gamma_{K_{11}-4U} (m_0)\;
\int \dd\gamma_{M_{11}} (\vec m)\;
\\
& \int \dd \gamma_{-D_{22}} (\bar \Delta, \Delta)\;
F(m_0,\vec m, \bar \Delta, \Delta)
\end{split}
\eeq
Here we have assumed that the signs of $D$, $M$, and $K-4U$ are such that the integrals converge. This is the case for $D$ and $M$. A more detailed discussion of $K-4U$ follows below. --
This transformation makes the action quadratic in the fermionic fields, so the integral over these fields can be performed, resulting in a Pfaffian and the exponential of a quadratic form in the fermionic source fields. After a few transformations, one arrives at
\beq\label{ZHHfinal}
\tilde Z (\bar H, H)
=
\left\langle 
\E^{
\frac12 \tr \log \bQ}\;
\E^{
\frac12\, (-\bar H, T_\Omega H) \; \bQ^{-1} 
\left(
{\begin{array}{c}
\scriptstyle T_\Omega H \\ \scriptstyle \bar H
\end{array}}
\right)
}
\right\rangle_{K,M,D}
\eeq
with $\bQ = \bI - \bL$, where $\bI$ is the identity operator, $\bI_{\alpha,\alpha'} (k,k') = \delta_{\alpha,\alpha'} \delta (k,k')$, and $\bL (k,k')$ is given as the product
\beq
\bL (k,k')
=
\left[
\begin{array}{cc}
T_\Omega (k)  
&
0 
\\
0 
&
1
\end{array}
\right]
\; 
\bM (k,k')
\;
\left[
\begin{array}{cc}
1
&
0 
\\
0 
&
T_\Omega (k')^t
\end{array}
\right]
\eeq
(where $^t$ denotes the transpose in the spin indices)
and 
\beq
\bM (k,k')
=
\left[
\begin{array}{cc}
\slashed{m} (k-k') 
&
2 \varepsilon f(\frac{k-k'}{2}) \Delta(k+k') 
\\
- 2 \varepsilon f(\frac{k-k'}{2}) \overline{\Delta(k+k')} 
&
\left(\slashed{m} (k'-k) \right)^t
\end{array}
\right]
\eeq
with $\slashed{m} = m_0 1_2 + \vec m \cdot \vec \sigma$ and $\varepsilon = \I \sigma_2$. 

The $\tr \log \bQ$ contributes a term to the bosonic action which can be expanded in the standard way as a sum over fermion loops with external boson lines. Thus it contributes linear terms in the boson fields, quadratic terms that modify the boson propagators, and higher order interaction terms. In general, these terms mix the different Bose fields, and this mixing can be rather nontrivial if coexistence of different phases is possible. 
The $\bQ^{-1}$ in the quadratic exponent involving the fermionic source fields has the interpretation of a fermion propagator in the background of the Bose fields. General fermionic $2n$-point functions are then obtained as derivatives with respect to the sources, hence as averages $\langle \cdot \rangle_{K,M,D}$ of Pfaffians of matrices with entries $\bQ^{-1} (k_i, k_j)$. 

The interacting bosonic field theory with the fields $m_0,\vec m, \Delta$ is the low-energy theory for the Hubbard model obtained from the RG flow at scale $\Omega$. The model still has the symmetries of the original action, but it is formulated in terms of fields coupling to the natural order parameters.

\section{Magnetism and Superconductivity} 
\label{sec:frgmfHubb}

\noindent
In the following, we specialize our analysis to mean-field theory, by keeping only the simplest configurations in the remaining functional integral.
Proceeding in this way is not the only way fRG and mean-field calculations can be combined. In particular one can use the method proposed in \cite{Wang2014}, which solves reduced models exactly. We choose the method presented in this section as a first step towards a more detailed analysis of the functional integral (\ref{ZHHfinal}). It allows us to continue previous fRG studies of \cite{UBHD-67255998,MS-GiSa} in a straightforward manner and it also provides a simple way of including the frequency dependent self-energy in the mean-field calculations.

We start by investigating the ferromagnetic and superconducting order parameters close to the QCP. We will test for gap formation in the vicinity of the critical hopping parameter at Van Hove filling, taking into account only the dominant part of the interaction, namely $M_{1,1}(0)$ responsible for a ferromagnetic interaction and $D_{2,2}(0)$ defining Cooper pair attraction. We drop the density-density term in the interaction before performing the HS transformation. Thus the transformation involves only $\vec m$ and $\Delta$ fields. The mean-field equations are the saddle-point equations for the bosonic effective action.  For the order parameters we consider, the mean-field solution can be assumed as constant in space and Euclidian time. Thus the functional integral is reduced to an integral over the zero modes of the two fields.
 
In this special case the Gaussian identities reduce to 
\begin{align}
	\E^{a^2/4} &= \frac{1}{\sqrt\pi} \int_{\mathbb R} \E^{-\phi^2 + \phi a} \dd{\phi} \label{eq:HS-real} \;, \\
	\E^{ab} &= \int_{\mathbb C} \E^{-|\phi|^2 + a \phi + b \overline \phi} \tfrac{\dd{\phi}\wedge\dd{\overline \phi}}{2\pi\I} \;. \label{eq:HS-complex}
\end{align}
After integrating out the fermions we obtain,
\begin{equation}
    Z \propto \int \mathrm{d}\Delta_\text{FM}
    \E^{ -\bV F_\text{FM}
    }\;,
\end{equation}
For the ferromagnetic ansatz
\begin{equation}
    F_\text{FM} = \frac{1}{M_{1,1}(0)} \Delta_\text{FM}^2
    \;-\sum\limits_{\sigma\in\{+,-\}} \int\limits_p \ln( 1 + \sigma \Delta_\text{FM} T(p) ) \;.
\end{equation}
We proceed similarly in the Cooper channel. A complex HS transformation yields,
\begin{equation}
    Z \propto \int \mathrm{d}\Delta_\text{dSC}\wedge \mathrm{d}\overline \Delta_\text{dSC}
    \E^{ - \bV F_\text{dSC}
    }\;,
\end{equation}
with
\begin{align}
    F_\text{dSC} =& \frac{1}{D_{2,2}(0)} |\Delta_\text{dSC}|^2\\
    &-\int\limits_p \ln( 1 +  |\Delta_\text{dSC}|^2 f_2^2(\vc k) T(p)T(-p) ) \;.
\end{align}
$F_\text{dSC}$ and $\;F_\text{FM}$ play the role of the free-energy per degree of freedom relative to the free-energy of the paramagnetic phase. In the thermodynamic limit and at zero temperature the saddle point with the dominant exponent (smallest free-energy) determines the phase of the system.
As already mentioned, coexistence of the two orders has been ruled out in this situation, as it corresponds to a maximum of the free energy.

$T(k)$ is given in Eq.~\eqref{eq:def-covar-T}. The self-energy entering the equations is the self-energy at scale $\Omega$ which in the vicinity of the critical hopping and at Van Hove filling, can be parametrized as \footnote{The difference in minus sign lies in different definitions of self-energy compared to \onlinecite{MS-GiSa}.}
\begin{equation} \label{eq:self-energy-fit}
    \Im \Sigma_\Omega(\omega)/\omega = -\frac{a}{(1+b^2 \omega^2)^{\gamma/2}}\;,
\end{equation}
where $a,b$ and $\gamma$ depend on the hopping amplitudes and the scale parameter \cite{MS-GiSa}. The right-hand side approaches a constant approximately as the frequencies drop below the stopping scale.
Within the static mean-field approximation we cannot calculate the small frequency behavior of the self-energy and incorporate it. To compensate for the loss we can use the extrapolation of the fRG data. By assuming that the extrapolation is the correct asymptotic behavior, we compute the gaps also using
\begin{equation} \label{eq:self-energy-fit-extrapolation}
    \Im \Sigma_\Omega^\text{ext.}(\omega)/\omega = -a(b^2 \omega^2)^{-\gamma/2}\;.
\end{equation}
At small stopping scales the difference we see in the order parameter using either version becomes negligible. If we remove the regulator we recover the toy-model of Section \ref{sec:mf-toy-model} (c.f.\ Eq.~\eqref{eq:self-energy-fit-extrapolation} and Eq.~\eqref{eq:toy-model-se}).

The saddle point conditions are
\begin{subequations}
\begin{align}
    &\Delta_\text{FM} = \frac{M_{1,1}(0)}{2} \intsp{k} \frac{\sigma}{T^{-1}(k)+\sigma \Delta_\text{FM}}\;, \label{eq:frgmf-fm}\\
    &1 =  D_{2,2}(0) \intsp{k} \frac{f_2^2(\vc k)}{T^{-1}(k)T^{-1}(-k)+f_2^2(\vc k)\abs{\Delta_\text{dSC}}^2}\;,  \label{eq:frgmf-dsc}
\end{align}
\end{subequations}
and the order parameters $O_\text{dSC}=2 \Delta_\text{dSC}/D_{2,2}(0), \; O_\text{FM}=2 \Delta_\text{FM}/M_{1,1}(0)$ are the expectation values of the bilinears
\begin{align}
    \widehat O_{\text{FM}} &= \intsp{k} \sigma\,{\overline\Psi}_{k,\sigma}\Psi_{k,\sigma}\;,\\
    \widehat O_{\text{dSC}} &= \frac12 \intsp{k} \sigma\,f_2(\vc k){\overline\Psi}_{k,\sigma}{\overline\Psi}_{-k,-\sigma}\;.
\end{align}

\begin{figure}[]
    \centering
    \includegraphics{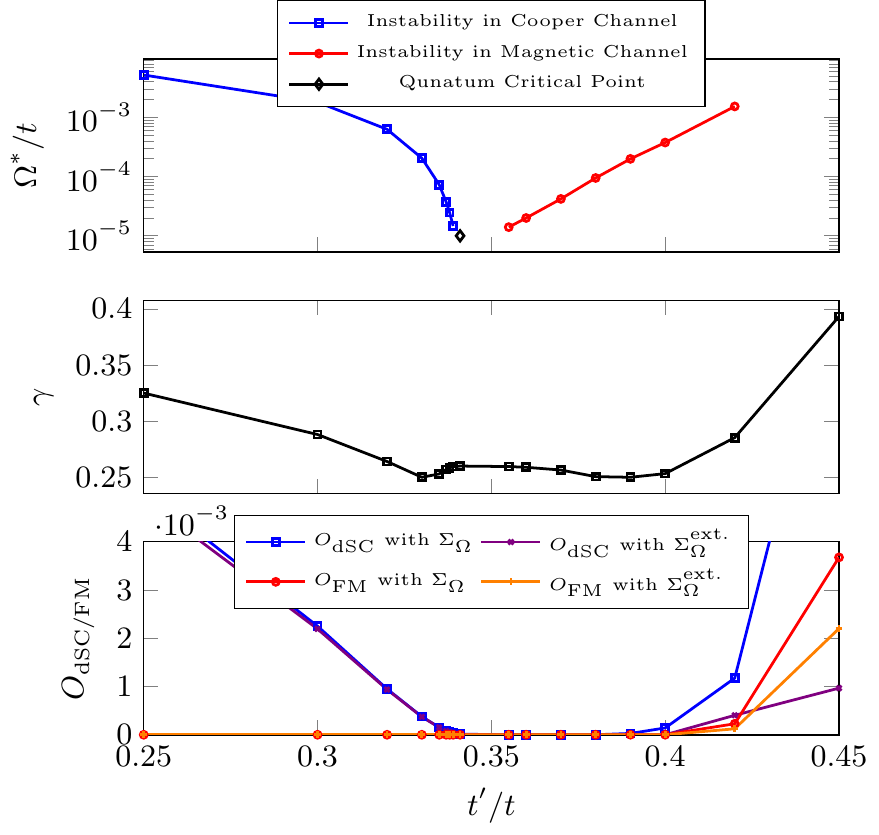}
   \caption{{The stopping scale $\Omega^*$ (taken from Refs.\ \onlinecite{MS-HuGiSa} and \onlinecite{MS-GiSa}), the anomalous exponent of the self-energy $\gamma$ (K.\ Giering, private communication), and the order parameters $O_\text{dSC},\;O_\text{FM}$ as functions of hopping parameter $t'/t$. In the range $(0.34,0.38)$ the order parameters vanish.}}   
%
%
    \label{fig:mf-rg}
\end{figure}

For the numerical calculations we use the fRG data given in Table \ref{table:fRGData}. The results are from previous studies in collaboration with K.~Giering and C.~Husemann, see Refs.\  \onlinecite{UBHD-67255998}, \onlinecite{MS-GiSa},  and \onlinecite{MS-HuGiSa} and the table caption. We use the conventions of Ref.\ \onlinecite{HuSa}, which differ by a factor of two in the definition of the bosonic propagators from the other references, and we have adapted the values in the table to this convention. The corresponding order parameters are shown in Figure \ref{fig:mf-rg}. In the range $t'/t \in (0.34,0.38)$, \pr{where $\gamma \mathop{\mathsmaller{\lessapprox}} 0.26$,} both order parameters vanish within numerical tolerance which confirms the existence of a QCP in this parameter region. The self-energy suppresses order in two ways: in the RG flow, it pushes the growth of the pairing interaction to very small scales, and in the mean-field equation it prevents gap formation in the way exemplified in the toy model of Section \ref{sec:mf-toy-model}.

\begin{table}[t]
\centering{
\small
\begin{tabular}{@{}|L|L|L|L|L|L|L|@{}}
\hline
t'/t & \log_{10} \Omega^*/t & a & \log_{10}b & \gamma & D_{2,2}(0)/t & M_{1,1}(0)/t \\ \hline
0.250  & -2.28  & 1.19  & 1.83 & 0.325 &  20.0 & 4.18 \\
0.300  & -2.74  & 1.60  & 2.34 & 0.288 &  20.0 & 5.75 \\
0.320  & -3.20  & 2.17  & 2.89 & 0.264 &  20.0 & 7.44 \\
0.330  & -3.69  & 2.93  & 3.42 & 0.250 &  20.0 & 9.54 \\
0.335  & -4.14  & 4.30  & 4.20 & 0.253 &  20.0 & 11.8 \\
0.337  & -4.43  & 5.49  & 4.49 & 0.257 &  20.0 & 13.3 \\
0.338  & -4.61  & 5.91  & 4.56 & 0.258 &  20.0 & 14.4 \\
0.339  & -4.83  & 6.24  & 4.61 & 0.259 &  20.0 & 15.7 \\
0.341  & -5     & 6.4   & 4.66 & 0.260 &  2.05 & 16.0 \\
0.355  & -4.85  & 6.27  & 4.62 & 0.260 &  0.05 & 20.0 \\
0.360  & -4.70  & 6.07  & 4.59 & 0.259 &  0.06 & 20.0 \\
0.370  & -4.38  & 5.33  & 4.47 & 0.256 &  0.09 & 20.0 \\
0.380  & -4.02  & 3.57  & 3.81 & 0.251 &  0.17 & 20.0 \\
0.390  & -3.70  & 2.95  & 3.43 & 0.250 &  0.26 & 20.0 \\
0.400  & -3.42  & 2.42  & 3.04 & 0.253 &  0.36 & 20.0 \\
0.420  & -2.82  & 1.67  & 2.40 & 0.285 &  0.55 & 20.0 \\
0.450  & -1.95  & 0.88  & 1.09 & 0.393 &  0.64 & 20.0 \\
\hline
\end{tabular}}
\caption{Compilation of fRG data (K.U.~Giering, Ref.\ \onlinecite{UBHD-67255998} and private communication. For definitions and details about the exponents $a$, $b$, and $\gamma$, see also Ref. \onlinecite{MS-GiSa}.). We have independently recalculated the values of $\Omega^*$, $D_{2,2}(0)$, and $M_{1,1}(0)$ at all given values of $t'/t$.} \label{table:fRGData}
\end{table}

\section{Mean-Field Effects of the Density-Density Interaction} \label{sec:mf-dd}
\begin{figure}[]
    {\includegraphics{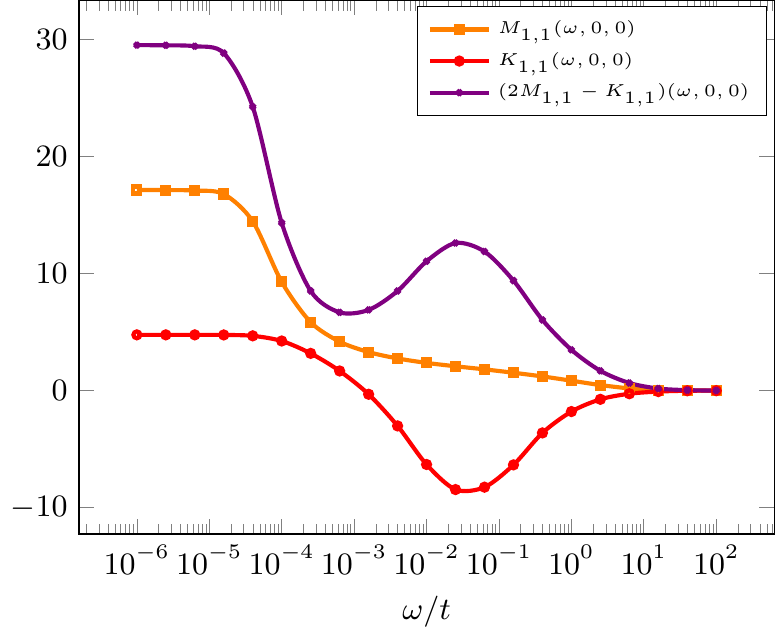}}
  \caption{Effective interaction in magnetic and scattering channel at the scale $\Omega/t=3\times 10^{-5}$ for the bare interaction $U/t=3$ at Van Hove filling and the critical hopping amplitude. Combining two channels, the overall density-density interaction $U+(2 M_{1,1}-K_{1,1})(\omega,0,0)$ is peaked at some nonzero frequency. In contrast to the data used in previous section, the ones shown here are from calculations that take the momentum dependence of self-energy into account.}
 \label{fig:density-density-interaction}
\end{figure}

\noindent
In Section \ref{sec:frgmf} we showed that in the vicinity of the QCP, magnetic ordering and Cooper pairing both become negligible. In this section we assume that we are in this quantum critical regime and investigate the effects of the density-density interaction on the propagator. As shown in Figure \ref{fig:density-density-interaction}, during the flow the effective interaction develops a strong peak in the scattering channel at a nonzero frequency. In the simplest approximation we project the effective interaction to a sum of delta distributions at zero frequency and at frequencies $\pm \tilde\omega \ne  0$. The zero frequency part will give rise to a Hartree self-energy, which we compensate by appropriate choice of the chemical potential to fix the density at Van Hove filling.

\subsection{Introduction} \label{subsec:notes-on-d-d-mf}
\noindent
In the density-density channel, depending on the sign of the interaction, the bosonic action obtained from a HST according to Eq.~\eqref{eq:HS-real} may not have a saddle point on the real axis. Provided that the action is holomorphic in the HS field we may be able to find a saddle point in the complex plane and deform the integration contour to pass through the saddle point without changing the result. If this deformation is possible, the zero temperature asymptotics of the partition and correlation functions would then contain an integral of the type
\begin{equation}
    F(\lambda) = \int_\gamma f(z) \E^{\lambda S(z)} \dd{z} \;.
\end{equation}
where $\gamma $ is the new contour. If $S$ has a single simple saddle point at an interior point $z_0$ of the integration contour $\gamma$, then the $\lambda \to \infty$ asymptotic of $F(\lambda)$ is given by
\begin{equation} \label{eq:mf-lemma-dd}
    F(\lambda) = \sqrt{\frac{2\pi}{-S''(z_0)}} \lambda^{-1/2} \E^{\lambda S(z_0)} \qty(f(z_0)+\mathcal{O}\qty(\lambda^{-1}))\;.
\end{equation}
For our purpose a constant shift of the integration contour is sufficient to make sure that the contour passes through the saddle point of the exponent. In particular we can formulate
the HST \eqref{eq:HS-real} in the more general form
\begin{equation} \label{eq:HS-real-ext}
	\E^{a^2/4} = \frac{1}{\sqrt\pi} \int_{\mathbb R} \E^{-(\phi+\I c)^2 + (\phi+\I c) a} \dd{\phi}  \;,
\end{equation}
where $c$ is an arbitrary real number.

In the case of a complex HST the generalization is as follows,
\begin{equation} \label{eq:HS-complex-ext}
\E^{ab} = \int_{\mathbb C} \E^{-(\phi+\I \psi)(\overline\phi+\I \overline \psi)  + a (\phi+\I \psi) + b (\overline \phi+\I \overline \psi)} \tfrac{\dd{\phi}\wedge\dd{\overline \phi}}{2\pi\I} \;,
\end{equation}
where $\psi\in \mathbb{C}$ is an arbitrary complex number. The proof is simple, let
$\phi=u+\I v,\; \psi=x+\I y$ with $u,v,x,y\in\mathbb{R}$, then
\begin{equation}
\begin{split}
I&=\int_{\mathbb C} \E^{-(\phi+\I \psi)(\overline\phi+\I \overline \psi)  + a (\phi+\I \psi) + b (\overline \phi+\I \overline \psi)} \tfrac{\dd{\phi}\wedge\dd{\overline \phi}}{2\pi\I}  \\&= \int_{\mathbb R^2} \E^{-(u+\I x)^2-(v+\I y)^2 + (a+b) (u+\I x) + \I(a-b)(v+\I y)} \tfrac{\dd{u}\wedge\dd{v}}{\pi} \;,
\end{split}
\end{equation}
Eq.~\eqref{eq:HS-complex-ext} now follows from Eq.~\eqref{eq:HS-real-ext},
\begin{equation}
I=\E^{(a+b)^2/4-(a+b)^2/4}=\E^{a b}\;.
\end{equation}
As long as the integrals in Eq.~\eqref{eq:HS-real-ext} and Eq.~\eqref{eq:HS-complex-ext} are considered exactly they are independent of $c$ and $\psi$, but only for suitable values of these quantities ---let us denote them by $c'$ and $\psi'$--- can the integrals be evaluated in saddle point approximation. Through the analytic structure of the interand, the integral then depends on these specific values $c'$ and $\psi'$.

\subsection{Mean-field equation for a frequency- and momentum-dependent density-density interaction}
\noindent
We consider a density-density interaction of the form
\footnote{here $K$ is a general density-density interaction, not necessarily equal to $K_{1,1}$.}
\begin{equation} \label{eq:MF:frmionic-action}
	\begin{split}
		S(\widetilde \psi,\psi) =& - \int_{\omega \vc{p} s} \overline\psi_{\omega \vc{p} s} (\I \omega - \epsilon_{\vc{p}}+\mu) \psi_{\omega \vc{p} s}\\
		&+ \frac14 \int_{p} K(p) S^{(0)}(p) S^{(0)}(-p)\;,
	\end{split}
\end{equation}
which does not depend on the non-transfer frequencies and momenta. 
\neuneu{$\int_{\omega \vc{p} s}$ represents a sum over Matsubara frequencies, momenta and spin with appropriate normalization,
\begin{equation}
    \int_{\omega \vc{p} s} \bullet = 
    \frac{1}{V \beta}\sum_{\omega\in \mathbb{M_\text{F}}/\mathbb{M_\text{B}}}\sum_{\vc p \in \Gamma^*}\sum_{s\in\{+,-\}}\bullet
\end{equation}
such that in the thermodynamic limit,
\begin{equation}
    \int_{\omega \vc{p} s} \to \frac{1}{(2\pi)^3} \sum_{s\in\{+,-\}} \int_{-\infty}^{\infty}\int_{[-\pi \pi)^2}\bullet \dd{\vc p} \dd{\omega}
\end{equation}
$\Gamma^*=(\frac{2\pi}{L}\mathbb{Z})^2/(2\pi \mathbb{Z})^2$ is the momentum space and $V=L^2$. 
The discretized Euclidean time axis is given by $\mathbb{T}_n:=\{-\beta/2+\beta k/n: k\in\{0,\dots,n-1\}\}$ and the corresponding fermionic and bosonic Matsubara frequencies are
\begin{equation}
\begin{split}
 \mathbb{M}_\text{F}&=\{\frac{\pi}{\beta}k : k \in (2\mathbb{Z}-1)\cap [-n,n)\}  \\ 
 \mathbb{M}_\text{B}&=\{\frac{\pi}{\beta}k : k \in (2\mathbb{Z})\cap [-n,n)\}\;.
 \end{split}
\end{equation}
In the following it is clear from the context whether the sum has to be taken over $\mathbb{M}_\text{F}$ or over $\mathbb{M}_\text{B}$.
}

In the following we assume $K$ to be real and symmetric with respect to $\omega\to -\omega$.
For the bare Hubbard model $K$ would be positive and constant.
Next we write the interaction term from Eq.~\eqref{eq:MF:frmionic-action} as
\begin{equation}
	\begin{split}
&- \frac14 \int_{p} K(p)S^{(0)}(p) S^{(0)}(-p) =\\
&-\frac{1}{4}\frac{1}{\beta V} K(0)S^{(0)}(0)^2 -\frac{1}{2} \frac{1}{\beta V} \sum_{p>0}K(p) S^{(0)}(p)  S^{(0)}(-p)\;,
\end{split}
\end{equation}
where ``$>$'' denotes lexicographical order on the product space $\mathbb{M}_\text{B} \times \Gamma^*$. 
Making the reflection symmetry explicit is not strictly necessary but it reduces the number of HS fields we need to define. Proceeding according to Eq.~(\ref{eq:HS-real-ext},\ref{eq:HS-complex-ext}) we obtain the mixed action
\begin{equation}
\begin{split}
&S_{\text{H.S.}} =
-\frac{1}{\bV} \sum_{\omega \vc{p} s} \widetilde\psi_{\omega \vc{p} s} (\I \omega - \epsilon_{\vc{p}}+\mu) \psi_{\omega \vc{p} s}\\
&+\bV \frac{\Phi_0^2}{|K(0)|} -\I I_K(0) \Phi_0 S^{(0)}(0)
+2 \bV \sum_{p>0} \frac{\Phi(p)\widetilde\Phi(p)}{|K(p)|}\\
& - \I \sum_{p>0} \left( I_K(p)S^{(0)}(p)\Phi(p) +I_K(-p)S^{(0)}(-p)\widetilde\Phi(p)\right)\;,
\end{split}
\end{equation}
where, for $f(x)\in \mathbb{R}$,
$I_f(x) = 1 \qif f(x)\ge 0$ and $I_f(x) =\I$ otherwise.

The field $\Phi$ includes a possible shift of the integration contour. To be precise $\Phi_0 = \phi_0+\I \psi_0$ with $\phi_0,\psi_0\in\mathbb{R}$ and for $p>0$, $\Phi(p)=\phi(p)+\I \psi(p),\widetilde \Phi(p)=\overline\phi(p)+\I \overline\psi(p)$ with $\phi(p),\psi(p)\in\mathbb{C}$. $\phi$ is the HS field and the field $\psi$ has to be chosen such that the integration contours pass through saddle points of the action.

Integrating out the fermions leads to the free energy
\begin{equation} \label{eq:dd-free-energy}
\begin{split}
F =
& \frac{\Phi_0^2}{|K(0)|}
+2 \sum_{p>0} \frac{\Phi(p)\widetilde\Phi(p)}{|K(p)|} \\
& - \frac{2}{\bV}\ln \det \Bigg [ (\I k_0 - \epsilon_{\vc{p}}+\mu+\I I_K(0)\Phi_0)\delta_{k,k'}  \\
&- \I\big(I_K(p-p')\Phi(p-p')\Theta(p-p'>0)\\
&+
I_K(p'-p)\widetilde\Phi(p'-p)\Theta(p'-p>0) \big) \Bigg]_{k,k'}\\
&+ C\;,
\end{split}
\end{equation}
where $C=\frac{2}{\bV}\ln \det [ (\I k_0 - \epsilon_{\vc{p}}+\mu)\delta_{k,k'} ]_{k,k'}$ is a normalziation constant.

In the general case a computation of the determinant is not feasible. The usual ansatz is to keep only the static part of the bosonic field. We will go a step further and take into account the field at some nonzero frequency $\tilde\omega$. The determinant can then be computed efficiently using the lemma presented in Appendix \ref{sec:lemma}.

\subsection{Numerical Setup}
\noindent
We make the following ansatz for $\Phi$,
\begin{equation}
	\label{eq:MF-numerical-results-ansatz-phi}
	\begin{split}
	\Phi(p) &=   \Phi_{\tilde \omega} \delta_{k_0,\tilde \omega} \delta_{\vc k,0}\;,\\
	\widetilde \Phi(p) &=  \widetilde\Phi_{\tilde \omega} \delta_{k_0,\tilde \omega} \delta_{\vc k,0}\;.
	\end{split}
\end{equation}
So in addition to the static field $\Phi_0$ at zero frequency the free energy depends on $\Phi_{\tilde \omega}$ and $\widetilde\Phi_{\tilde \omega}$, which incorporates a dependence on $K$ at the nonzero frequency $\tilde\omega$.
The free energy \eqref{eq:dd-free-energy} now reads
\begin{equation} \label{eq:mf-densiy-free-energy-after-ansatz}
\begin{split}
F =&\frac{\Phi_0^2}{|K(0)|}
+2 \frac{\Phi_{\tilde \omega}\widetilde\Phi_{\tilde \omega}}{|K(\tilde \omega)|}
\\
&- \frac{2}{\bV}\sum_{\vc k} \ln \det \Bigg [ (\I k_0 - \epsilon_{\vc{p}}+\mu+\I I_K(0)\Phi_0)\delta_{k_0,k'_0}  \\
&\quad\quad\quad-\I I_K(\tilde \omega) \left(\Phi_{\tilde \omega}\, \delta_{k_0-k'_0,\tilde \omega} + \widetilde\Phi_{\tilde \omega}\, \delta_{k'_0-k_0,\tilde \omega} \right) \Bigg]_{k_0,k'_0}\\& + C\;.
\end{split}
\end{equation}
Independent of whether $K(0)$ is positive or negative we can adjust the chemical potential to ensure Van Hove filling. Then at the saddle point of $\Phi_0$ we have $\mu+\I I_K(0)\Phi_0=0$. At fixed density we only need to find the saddle point of $F$ as a function of $\Phi_{\tilde \omega}$ and $\widetilde\Phi_{\tilde \omega}$. For fixed $\Phi_0$ and at Van Hove Filling, $F$ is then up a constant given by the $n\to \infty$ limit of
\begin{equation}
\begin{split}
F_n =&2 \frac{\Phi_{\tilde \omega}\widetilde\Phi_{\tilde \omega}}{|K(\tilde \omega)|}
- \frac{2}{\beta}\int_{\vc k} \ln \det \Bigg [ (\I \dfreq\omega_k - \epsilon_{\vc{p}}
)\delta_{k,k'}\\
&-\I I_K(\tilde \omega) \left(\Phi_{\tilde \omega}\, \delta_{k-k',m} + \widetilde\Phi_{\tilde \omega}\, \delta_{k'-k,m} \right) \Bigg]_{k,k'} 
\end{split}
\end{equation}
where
\begin{equation}
	\label{eq:hat-omega}
	\dfreq \omega_k = -\I \frac{n}{\beta}\qty(1-\E^{-\I \pi (2 k-1)/n})
\end{equation}
and $\tilde\omega = \frac{2\pi}{\beta}m$, for some fixed $m\in\{1,\dots,n-1\}$.

 $F_n$ depends on the HS fields $\Phi_{\tilde \omega}$ and $\widetilde\Phi_{\tilde \omega}$ only through their product. For the determinant this follows from Eq.~\eqref{eq:lemmad-det:recurrent}. As a result any non-trivial saddle point of $F$ as a function of $\Phi_{\tilde \omega}$ or $\widetilde\Phi_{\tilde \omega}$ is a saddle point of $F$ as a function of $\Phi_{\tilde \omega}\widetilde\Phi_{\tilde \omega}$,
\begin{equation}
	\pdv{F}{\Phi_{\tilde \omega}} = \widetilde\Phi_{\tilde \omega} \pdv{F}{(\Phi_{\tilde \omega}\widetilde\Phi_{\tilde \omega})} \overset{!}{=} 0 \;.
\end{equation}

To summarize, we need to find the saddle point of $F_n(K(\tilde \omega),z)$ as a function of $z=\Phi_{\tilde \omega}\widetilde\Phi_{\tilde \omega}$ given through
\begin{alignat}{2}
	&F_n(\tilde K;z) &&= 2 \frac{z}{\abs{K(\tilde \omega)}} - f^{\mathrm{sgn(\tilde K)}}_n(z)\;, \notag\\
	&f^{\pm}_n(z) &&=\frac{2}{\beta} \int \dd{\epsilon} \rho(\epsilon) \ln f_n^{\pm}(z,\epsilon)\;,\notag\\
	&f_{-1}^{\pm}(z,\epsilon) &&= 0 \quad,\quad f_0^{\pm}(\epsilon) = 1 \;, \notag \\
	&f_m^{\pm}(z,\epsilon) &&= f_{m-1}^{\pm}(z,\epsilon) \\
& && \phantom{=}\pm z \frac{ \delta_{\abs{[i]_{n,k}-[j]_{n,k}},m}\,f_{m-2}^{\pm}(z,\epsilon)}{\qty(\I \dfreq\omega(2[i]_{n,k})-\epsilon)\qty(\I \dfreq\omega(2[j]_{n,k})-\epsilon)} \;, \label{eq:mf-density:full-eqs}
\end{alignat}
where $\rho(e) = \frac{1}{(2\pi)^2}\int_{[-\pi,\pi)^2}  \delta(\epsilon(\vc k)-e) \dd^2{\vc k}$ is the density of states.

The following symmetry considerations can greatly reduce the numerical effort.\\
First, since
\begin{equation}
	f^+_n(z) = f^-_n(-z)\;,
\end{equation}
if $z_0$ is a saddle point of $F_n(\tilde K,z)$ then $-z_0$ is a saddle point of $F_n(-\tilde K;z)$. So we may restrict ourselves to case $\tilde K=K(\tilde \omega)>0$ in the following.

Second, it easy to show that
\begin{equation}\label{eq:Fccsymmetric}
	\overline{F_n(\tilde K,z)} = F_n(\tilde K,\overline z) \;.
\end{equation}
Thus, for fixed $\tilde K$, if $z\in\mathbb{R}$ then $F_n(\tilde K;z)$ is real.

Let $z=x+\I y$ and $f:z\mapsto F_n(\tilde K;z) = u(x,y)+\I v(x,y)$ with $x,y,u,v\in \mathbb{R}$. Let $z_0=x_0+\I\,0$ and $f$ be holomorphic at $z_0$ then if ${\pdv{u}{x}}(x_0,0)=0$ since ${\pdv{v}{x}}(x_0,0)=0$ (because of Eq.~\eqref{eq:Fccsymmetric}), it follows from Cauchy-Riemann equations that $f'(z_0)=0$, i.e.\ $z_0$ is a saddle point of $f$ if $x_0$ is a saddle point of $f|_\mathbb{R}$.

\subsection{The Quasiparticle Lifetime}
\noindent
In the following we fix $n=10^4$. We have verified numerically that our results are then stable and don't change much if $n$ is chosen to be even larger. 

\begin{figure}
\includegraphics{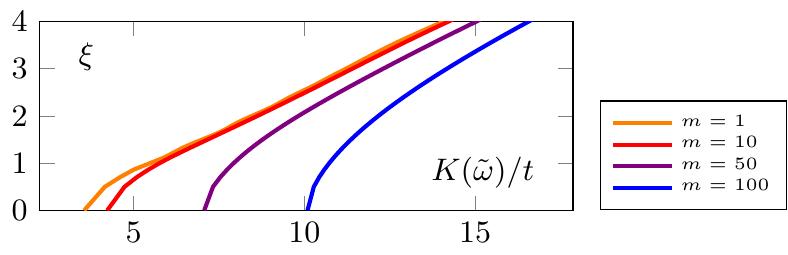}
\caption{Saddle point position in terms of $\xi=\sqrt{\abs{z}}$ as a function of interaction strength $K(\tilde\omega)$.}
\label{fig:MF-density:xiK}
\end{figure}

For large enough $K(\tilde \omega)$ the free energy has nontrivial saddle points.
In particular when $K(\tilde \omega)>0$ there is exactly one saddle point on an integration contour $\gamma:\mathbb{R}\mapsto \mathbb{C},t\mapsto t+\I a$, for a suitable value of $a$ which we denote by $a=\psi$. This saddle point lies on the imaginary axis. In our notation, at the saddle point, the real part of $\Phi=\phi+\I \psi$ is equal to zero and $z=\Phi \widetilde \Phi<0$. In the case $K(\tilde \omega)\leq 0$ there is one saddle point on the real axis and $z=\Phi \widetilde \Phi \geq 0$. Note that all measurable quantities should be independent of the sign of $K(\tilde \omega)$ for a nonzero frequency $\tilde \omega$ because of translational invariance along the Euclidean time axis and the periodicity of the interaction. We denote the saddle point of the free-energy for a given $K(\tilde \omega)$ by $\zeta = \sqrt{\abs{z}}$. The correspondence is shown in Figure \ref{fig:MF-density:xiK}.

In saddle point approximation the fermionic propagator is then given by the diagonal of
\begin{equation} \label{MF-density:propagator}
\begin{split}
&G =
\qty[ (\I k_0 - \epsilon_{\vc{p}})\delta_{k,k'}
+ \zeta \left(\delta_{k_0-k'_0,\tilde \omega} + \delta_{k'_0-k_0,\tilde \omega} \right) ]^{-1}_{k,k'} =\\
&
\qty[ \delta_{k_0,k'_0}
+ \zeta \frac{\left(\delta_{k_0-k'_0,\tilde \omega} + \delta_{k'_0-k_0,\tilde \omega} \right)}{\sqrt{\I k_0 - \epsilon_{\vc{p}}}\sqrt{\I k'_0 - \epsilon_{\vc{k'}}}}\delta_{\vc k,\vc k'} ]^{-1}_{k,k'}\\
&\quad\quad\quad\quad\quad\quad\quad\quad\quad \qty[\frac{\delta_{k,k'}}{\sqrt{\I k_0 - \epsilon_{\vc{p}}}\sqrt{\I k'_0 - \epsilon_{\vc{k'}}} }]_{k,k'} \;.
\end{split}
\end{equation}
As expected it does not depend on the sign of $K(\tilde \omega)$.
If we consider the propagator $G$ as a function of the sum and difference of the two involved frequencies, i.e.\ $\omega_{\pm}=k_0 \pm k'_0$, i.e.\ $G_{k,k'}=\delta_{\vc k,\vc k'}G_{\epsilon(\vc k)}(\omega_-,\omega_+)$. Then $G_{\epsilon}(\omega_-,\omega_+)$ is \pr{non-zero} only when $\omega_-$ is a multiple of $\tilde \omega$. For fixed $\epsilon$ and $\omega_+$, $G_{\epsilon}(\omega_-,\omega_+)$ decays as $\abs{\omega_-}$ grows.

\begin{figure}
 \centering
	{\includegraphics{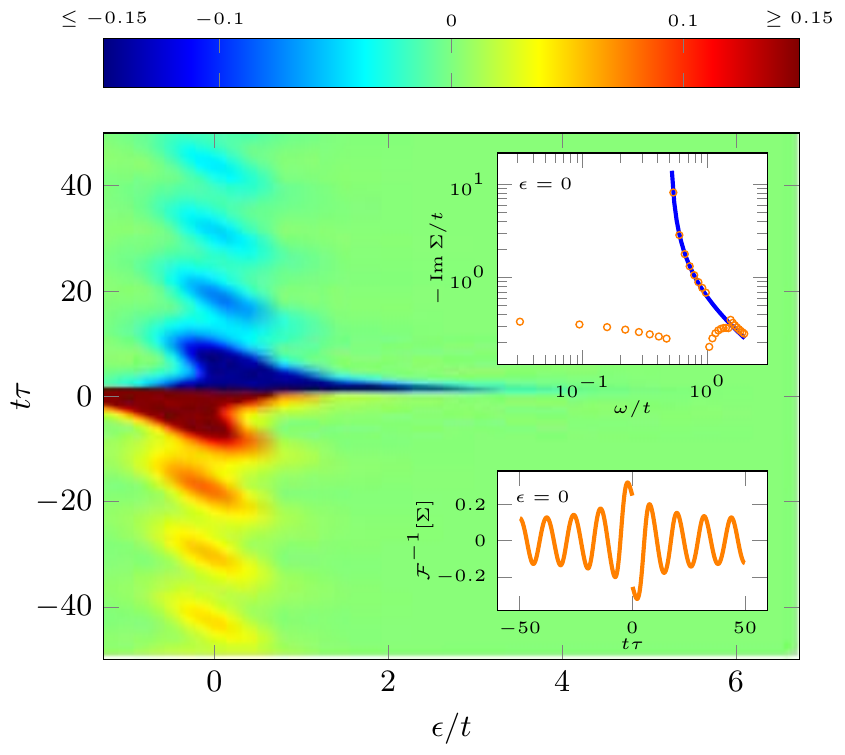}}

    \caption{Full propagator as a function of Euclidean time for $\zeta=0.5,m=8,n=10^4$. The periodic density-density interaction generated a pattern with $m$ peaks in the correlation function. The upper inset shows the imaginary part of the Dyson self-energy computed with $\zeta=0.5,m=8,n=10^4$. The data points in the range $\tilde\omega < \omega < 2 \tilde\omega$ define the most prominent structure in the self-energy. They can be fitted well by $a (\omega-\tilde \omega)^\alpha$ with an $\alpha\approx-1$ (blue curve). The lower inset shows the Fourier transform of the self-energy at $\epsilon=0$.
    }
	\label{fig:MF-density:full-prop}

\end{figure}

In this notation the fermionic propagator $g(k_0)=g_{\epsilon(\vc k)}(k_0)$ is given by $g_\epsilon(\omega) = G_{\epsilon}(0,2\omega)$. For finite values of $\zeta$ or equivalently corresponding interaction strength $K(\tilde \omega)$ the propagator develops some non-trivial structure most prominent around $\omega=\tilde \omega$. $\tilde \omega$ is itself not on the frequency lattice as it is a bosonic frequency. An example of the imaginary part of the Dyson self-energy $\Sigma=c^{-1}-g^{-1}$ with $c=(\I \dfreq \omega -\epsilon)^{-1}$ is shown in the upper inset of Figure \ref{fig:MF-density:full-prop}. The behavior of the self-energy changes at multiples of $\tilde\omega$. The strongest peak between $\tilde \omega$ and $2 \tilde \omega$ can be fitted well with a model $a(\omega-\tilde\omega)^\alpha$ with $\alpha\approx -1$. A pole at $\omega=\pm \tilde\omega$ would translate to a discontinuity in time domain,
\begin{equation}
	\begin{split}
	&\mathcal{F}^{-1}_\omega\qty[- \I \qty(\frac{1}{\omega-\tilde\omega}+\frac{1}{\omega+\tilde\omega})](\tau) \\
	&\quad\quad\quad\quad\quad\quad\quad\quad= -\sqrt(2 \pi) \cos(\tilde \omega \tau) \mathrm{sgn}(\tau) \;.
\end{split}
\end{equation}

The inverse Fourier transform of the free propagator to Euclidean time in the limit $\zeta\to 0$ is given by
\begin{equation} \label{eq:MF-density-prop-xi0}
\begin{split}
	&\frac{1}{\beta} \sum_{n} \E^{-\I \omega_n \tau} c_{\epsilon}(\omega) =\\
	&\quad\quad\quad\quad\quad\begin{cases}
	\E^{-\epsilon\tau}f_\beta(\epsilon) &\qif{-\beta<\tau\le 0} \\
	-\E^{-\epsilon\tau}(1-f_\beta(\epsilon)) &\qif{0<\tau<\beta}
	\end{cases} \;.
\end{split}
\end{equation}
\pr{In the interacting case with $\zeta=0.5$ and $m=8$ ($\tilde \omega/t\approx 0.5$) the result is shown in figure} \ref{fig:MF-density:full-prop}. The effect of the periodic density-density interaction $K(\tilde \omega)$ is evident in the correlation function which now shows $m$ distinct peaks. The inverse Fourier transform of the self-energy is shown in the lower inset of Figure \ref{fig:MF-density:full-prop}. It is remarkably simple compared to the self-energy in frequency space. Taking the limit $n\to \infty$ numerically it has a single discontinuity at $\tau=0$ in the interval $[-\beta,\beta)$.

 For small frequencies the imaginary part of the self-energy is approximately constant but discontinuous at the origin. Close to the Fermi surface and for small frequencies the full propagator behaves as
\begin{equation}
	g\approx \frac{1}{\I \dfreq \omega - \epsilon_{\vc k} + \I \text{sgn}(\omega)\tau_L^{-1} }
\end{equation}
where $\tau_L$ can be interpreted as the quasiparticle lifetime. 
In the limit $\zeta\to 0$ the quasiparticle lifetime diverges but as $\zeta$ or equivalently the interaction strength grows the quasiparticle lifetime becomes finite and drops rapidly. This is in full agreement with the assumption of criticality in this parameter regime.

\section{Conclusions}

\noindent
We have shown that in the repulsive 2D-Hubbard model, gap formation is suppressed in the vicinity of the critical hopping parameter ratio $\theta^\star=0.341$ and at Van Hove filling. The quantum-critical behavior is tightly related to the low-frequency asymptotics of the self-energy close to the Van Hove points. Previous fRG studies suggest a power-law behavior $\sim\text{sgn}(\omega)|\omega|^\alpha$ with $\alpha\approx 0.74$\cite{UBHD-67255998,MS-GiSa}. The fRG result is based on the extrapolation of the self-energy at some nonzero stopping scale. The absence of gaps in the range $t'/t\in(0.34,0.38)$, where the magnetic instability is quite pronounced, may be a consequence of neglecting the momentum dependence of the self-energy in the RG flow. We also want to point that chaining fRG and MF --as we did-- may lead to an underestimation of the gap parameters. However, by turning the self-energy on and off we can verify that the gaps are suppressed due to the self-energy effects rather than a low stopping scale of the flow. In further work, one may take the RG flow to the symmetry broken phase or perform mean-field calculations using the 2PI vertex extracted from fRG results \cite{Wang2014}. At this stage, beside quantitative results, we are also interested in the phenomenology of the physics in the vicinity of the QCP and the simplest methods serve well in this regard.

We have also used the mean-field approximation to calculate the influence of the frequency-dependent density-density interaction on the low-frequency structure of the self-energy. In general, the involved determinant is very challenging, so we investigated a minimal model. In this model we had a mono-frequency density-density interaction beside the static one. We found that such an interaction, which mimics the actual density-density interaction at the critical point, suppresses the quasiparticle lifetime.
Considering how much more complicated the actual interaction is one should suspect at least quantitative errors and treat these results with reservation, but exploring the minimal model has given us valuable information about the saddle-point structure of the free-energy which will be helpful for future analysis.

\begin{appendix}

\subsection{The Determinant of a Tridiagonal Matrix and its Generalization}
\label{sec:lemma}
\noindent
It is well known that the determinant $f_n = \det T_n$ of a tridiagonal matrix
\begin{equation}
	T_n = \begin{pmatrix}
	a_1 & b_1 \\
	c_1 & a_2 & b_2 \\
	& c_2 & \ddots & \ddots \\
	& & \ddots & \ddots & b_{n-1} \\
	& & & c_{n-1} & a_n
	\end{pmatrix}
\end{equation}
satisfied the recurrence relation
\begin{alignat}{2} \label{eq:tridiagonal:recurrent}
	&f_{-1} &&= 0 \;,\notag \\
	&f_0 &&= 1 \;,\notag \\
	&f_n &&= a_n f_{n-1} - c_{n-1}b_{n-1}f_{n-2}\;.
\end{alignat}
This relation can can be generalized as follows.
Let $T_n$ be a matrix of order $n\times n$ of the form
\begin{equation} \label{eq:MF-lemma-form}
	(T_n)_{ij} = \begin{cases}
 & a_i \qif i=j \\
 & b_i \qif j=i+k \\
 & c_j \qif i=j+k \\
 & 0   \qq{otherwise}
\end{cases}\;.
\end{equation}
For each $n$ let $d=\text{gcd}(k,n)$, $n'=n/d$ and $k'=k/d$. The map
\begin{equation}\label{eq:lemma-det:det}
	[m]_{n,k} := \Big\lfloor \frac{m-1}{n'} \Big\rfloor + \qty(k(m-1) \,\text{mod}\, n) + 1 \;
\end{equation}
defines a permutation which can be used to transform $T_n$ into a tridiagonal matrix. This transformation can be used to compute the determinant of $T_n$ efficiently. In particular if,
\begin{alignat}{2} \label{eq:lemmad-det:recurrent}
	&f_{-1} &&= 0 \;,\notag \\
	&f_0 &&= 1 \;,\notag \\
	&f_m &&= a_{[m]_{n,k}} f_{m-1} - c_{[m-1]_{n,k}}b_{[m-1]_{n,k}}f_{m-2}\;,
\end{alignat}
then $\det T_n = f_n$.

Proof: Every element $m \in \{1,2,\dots,n\}$ can be represented as $m = \ell n' + r + 1$ where $\ell = \Big\lfloor \frac{m-1}{n'} \Big\rfloor$ and $r=((m-1) \text{ mod } n')$. We define an isomorphism $m\mapsto [m]_{n,k} = (\ell + k r \text{ mod } n) + 1$. The matrix elements of $(T'_n)_{i,j}:=(T_n)_{[i]_{n,k},[j]_{n,k}}$ can only be nonzero if $((i-j)\text{ mod } n') \leq 1$. Thus $T'_n=\text{diag}(A_{0},A_{1},\dots,A_{d-1})$ is block diagonal and each block is of the form
\begin{equation}
A_{\ell} = \begin{pmatrix}
		\tilde a^{(\ell)}_{1} & \tilde b^{(\ell)}_{1} &&& \tilde c^{(\ell)}_{n'} \\
		\tilde c^{(\ell)}_{1} & \tilde a^{(\ell)}_{2} & \tilde b^{(\ell)}_{2} \\
		& \tilde c^{(\ell)}_{2} & \ddots & \ddots \\
		& & \ddots & \ddots & \tilde b^{(\ell)}_{n'-1} \\
		\tilde b^{(\ell)}_{n'} & & & \tilde c^{(\ell)}_{n'-1} & \tilde a^{(\ell)}_{n'}
\end{pmatrix}\;,
\end{equation}
where
\begin{equation}
	\begin{split}
		\tilde b^{(\ell)}_{n'} &= (T_n)_{[\ell n'+n']_{n,k},[\ell n'+1]_{n,k}} = (T_n)_{\ell+n-k+1,\ell+1} \;,\\
		\tilde c^{(\ell)}_{n'} &= (T_n)_{[\ell n'+1]_{n,k},[\ell n'+n']_{n,k}} = (T_n)_{\ell+1,\ell+n-k+1} \;.
	\end{split}
\end{equation}
Either $n\neq 2k$, then both $\tilde b^{(\ell)}_{n'}$ and $\tilde c^{(\ell)}_{n'}$ are zero or $n=2k$, then the blocks $A_\ell$ are $2\times 2$ matrices. In both cases $T'_n$ is tridiagonal and its determinant which is equal to the determinant of $T_n$ is given by Eq.~\eqref{eq:lemma-det:det}.

\end{appendix}

\end{document}